\documentclass[10pt,journal,compsoc]{IEEEtran}
\usepackage{amsmath,amssymb,amsfonts}
\usepackage{algorithmic}
\usepackage{graphicx}
\usepackage{textcomp}
\usepackage{xcolor}
\usepackage[linesnumbered,ruled]{algorithm2e}

\usepackage{booktabs}
\usepackage{multirow}
\usepackage{url}

\usepackage{bm}
\usepackage{enumitem}
\usepackage{subfigure}
\usepackage{amsthm}

\newcommand{\note}[1]{{\color{blue}{[\textbf{Attention}: #1]}}}
\newcommand{\revision}{{\color{black}}}

\ifCLASSOPTIONcompsoc
  \usepackage[nocompress]{cite}
\else
  \usepackage{cite}
\fi
\ifCLASSINFOpdf
\else
\fi
\begin{document}
%
\title{RPT: Toward Transferable Model on Heterogeneous Researcher Data via Pre-Training}
%
%
%
%

\author{{Ziyue Qiao, Yanjie Fu, Pengyang Wang, Meng Xiao, Zhiyuan Ning, Denghui Zhang, \\Yi Du, and Yuanchun Zhou}
\IEEEcompsocitemizethanks{\IEEEcompsocthanksitem Ziyue Qiao is with Computer Network Information Center, Chinese Academy of Sciences, and University of Chinese Academy of Sciences, Beijing.
E-mail: qiaoziyue@cnic.cn
\IEEEcompsocthanksitem Yanjie Fu is with Department of Computer Science, University of Central Florida, Orlando. E-mail: yanjie.fu@ucf.edu
\IEEEcompsocthanksitem Pengyang Wang is with Department of Computer and Information Science, University of Macau, Macau. E-mail: pywang@um.edu.mo
\IEEEcompsocthanksitem  Meng Xiao, Zhiyuan Ning, Yi Du, and Yuanchun Zhou are with Computer Network Information Center, Chinese Academy of Sciences, Beijing. E-mail: \{shaow, ningzhiyuan, yidu, yuanchunzhou\}@cnic.cn
\IEEEcompsocthanksitem  Denghui Zhang are with Information System department at Rutgers University, USA. E-mail: denghui.zhang@rutgers.edu
\IEEEcompsocthanksitem Corresponding author: Yi Du.
}
}

%
%

\markboth{Journal of \LaTeX\ Class Files,~Vol.~14, No.~8, August~2015}%
{Shell \MakeLowercase{\textit{et al.}}: Bare Demo of IEEEtran.cls for Computer Society Journals}
%



\IEEEtitleabstractindextext{%
\begin{abstract}
\revision{ With the growth of the academic engines, the mining and analysis acquisition of massive researcher data, such as collaborator recommendation and researcher retrieval, has become indispensable for improving the quality and intelligence of services.
However, most of the existing studies for researcher data mining focus on a single task for a particular application scenario and learning a task-specific model, which is usually unable to transfer to out-of-scope tasks.
In this paper, we propose a multi-task self-supervised learning-based researcher data pre-training model named RPT, which is efficient to accomplish multiple researcher data mining tasks.
Specifically, we divide the researchers' data into semantic document sets and community graph. We design the hierarchical Transformer and the local community encoder to capture information from the two categories of data, respectively. Then, we propose three self-supervised learning objectives to train the whole model.
For RPT's main task, we leverage contrastive learning to discriminate whether these captured two kinds of information belong to the same researcher. In addition, two auxiliary tasks, named hierarchical masked language model and community relation prediction for extracting semantic and community information, are integrated to improve pre-training. Finally, we also propose two transfer modes of RPT for fine-tuning in different scenarios.
We conduct extensive experiments to evaluate RPT, results on three downstream tasks verify the effectiveness of pre-training for researcher data mining.
}
\end{abstract}

\begin{IEEEkeywords}
Pre-Training, Contrastive Learning, Transformer, Graph Representation Learning
\end{IEEEkeywords}}

\maketitle

\IEEEdisplaynontitleabstractindextext

%
\IEEEpeerreviewmaketitle

\IEEEraisesectionheading{\section{Introduction}\label{sec:introduction}}
\IEEEPARstart{W}{ith} the pervasiveness of digital bibliographic search engines, e.g., Google Scholar, Aminer, Microsoft Academic Search, and DBLP, efforts have been dedicated to mining scientific data, one of the main focuses is researcher data mining, which aims to mine researchers' semantic attributes and community relationships for important applications, including: collaborator recommendation\cite{liu2018context, wang2019csteller}, academic network analysis\cite{tang2008arnetminer, dong2017metapath2vec, 9005458}, and expert finding\cite{zhang2020multi, dehghan2019translations}.

Millions of researchers and relevant data have been added to digital bibliographic datasets every year. The scientific research achievements and academic cooperation of researchers saw a continuation in the upward trend.
However, different researcher data mining tasks usually choose particular features and design unique models, and minimal works can be transferred to out-of-domain tasks. For example, previous studies \cite{yan2012scholarly} are more inclined to use graph representation learning-based models to explore the researcher community graphs in the collaborator recommendation task. In researchers' research field classification task, their publications are more valuable features, and the semantic models are more often used.
Hence, when multiple mining tasks on a tremendous amount of researcher data, it would be laborious for feature selection and computationally expensive to train different task-specific models, especially when these models need to be trained from scratch.
Also, most researcher data mining tasks need labeled data, which is usually quite expensive and time-consuming, especially involving manual effort. The deficiency of labeled data makes supervised models are easily over-fitting. And the unlabelled graph data is usually easily and cheaply collected.
Motivated by these, we aim to exploit the intrinsic data information disclosed by the unlabeled researcher data to train a generalized model for researcher data mining, which is transferable to various downstream tasks.

The pre-training technology is a proper solution, which has drawn increasing attention recently in many domains, such as natural language processing (NLP)\cite{kenton2019bert, yang2019xlnet}, computer vision (CV)\cite{he2020momentum, chen2020a, zhang2016deep}, and graph data mining\cite{qiu2020gcc, hu2020gpt}.
The idea is to first pre-train a general model to capture useful information on a large unlabeled dataset via self-supervised learning, and then, the pre-trained model is treated as a good initialization of the downstream tasks, further trained along with the downstream models for different application tasks with a few fine-tuning steps.
The pre-training model is a transferable model and convenient to train because the unlabeled data is easily available. In fine-tuning, the downstream models can be very lightweight.
Thus, this sharing mechanism of pre-training models is more efficient than independently training various task-specific models.
Early attempts to pre-training mainly focus on learning word or node representation, which optimizes the embedding vectors by preserving some similarity measure, such as the word co-occurrence frequency in texts and the network proximity in graphs, and directly uses the learned embeddings for downstream tasks. However, the embeddings are limited in preserving extensive information in large datasets. \revision{Thus, recent studies consider a model transfer setting that greatly improve the model ability of knowledge transferring~\cite{chen2020a, kenton2019bert, qiu2020gcc}. The schema is first pre-training a deep neural network-based encoder to incorporate the unsupervised information into the parameters, then fine-tuning it along with the the downstream model head (e.g, a classifier or a decoder) in an end-to-end manner.}




Inspired by these improvements, we aim to conduct pre-training on researcher data for mining tasks. The study in \cite{gururangan2020don} has proved the effectiveness of domain data for pre-training, and some work has leveraged pre-training models on academic domain data.
For example, SciBERT\cite{beltagy2019scibert} leverages BERT on scientific publications to improve the performance on downstream scientific NLP tasks.
SPECTER\cite{cohan2020specter} introduces the citation information into the pre-training to learn document-level embeddings of scientific papers.
However, existing pre-training models can not be directly applied to researcher data pre-training. Because the researcher data is more heterogeneous, including textual attributes (e.g., profiles, publications, patents, etc.) and graph-structured community relationships (e.g., collaborating, advisor-advisee, etc.). In contrast, most present models can only deal with a specific type of data.
Also, to extract information from the heterogeneous researcher data, the pre-training model should be more complex and heavy-weight than traditional ones, which brings new challenges to the pre-training of models on large-scale data.


\begin{figure}
  \centering
  \includegraphics[width=0.48\textwidth]{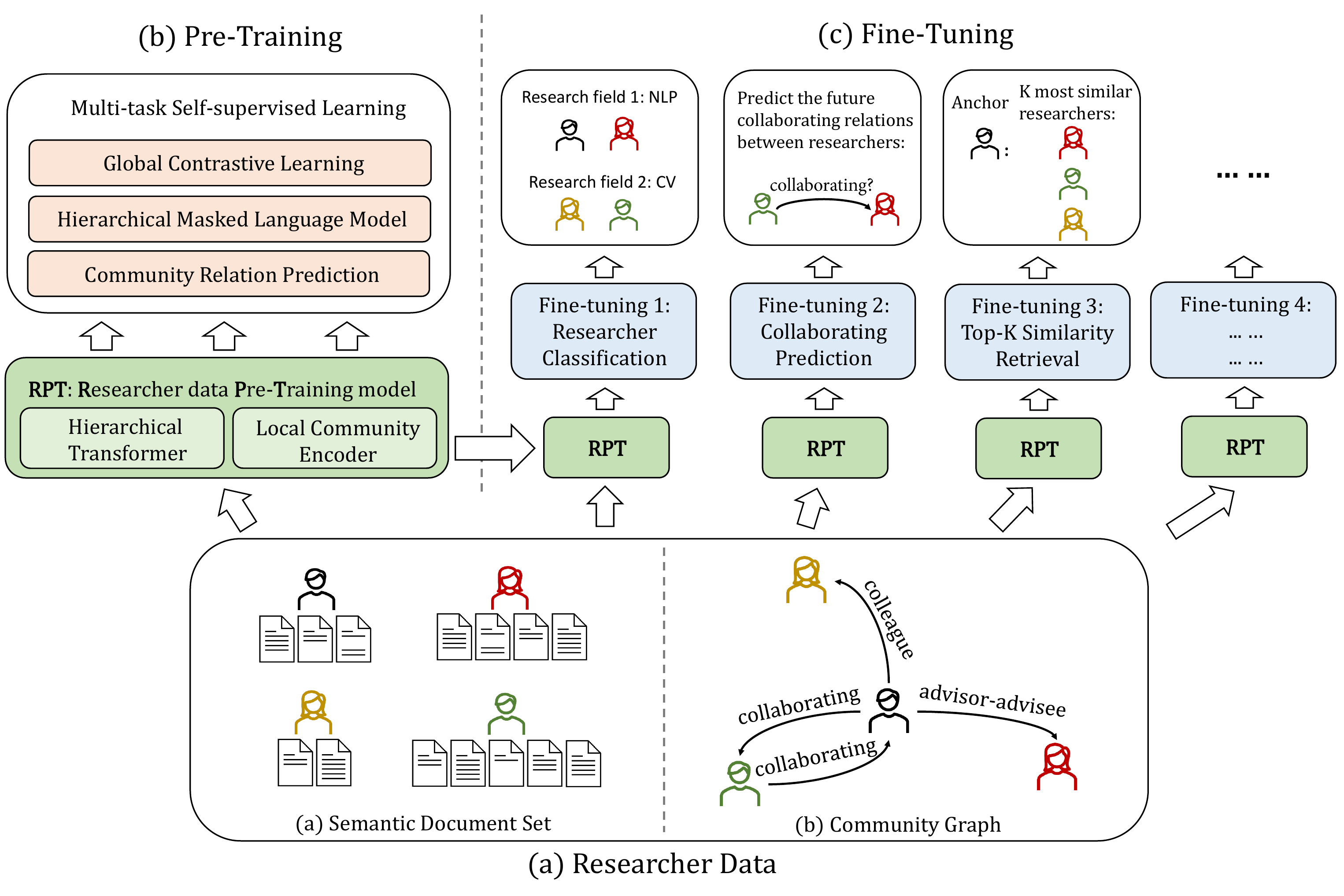}
  \caption{An illustration of our proposed pre-training and fine-tuning framework for researcher data mining. (a) We divide the researchers’ data into semantic document sets (containing researchers' textual attributes) and community graphs (preserving researchers' community relationships) as inputs for the RPT model and (b) pre-train the RPT model via a multi-task self-supervised learning objective to preserve the researcher data. (c) Then, we fine-tune the pre-trained RPT model via the objectives of different downstream tasks for researcher data mining.}
  \label{intro}
\end{figure}

In this paper, we leverage the idea of multi-task learning and self-supervised learning to design the \textbf{R}esearcher data \textbf{P}re-\textbf{T}raining(RPT) model. Specifically, for each researcher, we propose a hierarchical Transformer as the textual encoder to capture semantic information in researcher textual attributes. We also propose a linear local community sampling strategy and an efficient graph neural network (GNN) based local community encoder to capture the local community information of researchers. To leverage optimization on these two encoders, the main task of RPT is a contrastive learning model to discriminate whether these two captured information belongs to the same researcher. We also design two auxiliary tasks, Hierarchical Masked Language Model (HMLM) and Community Relation Prediction (CRP), respectively, to extract token-level semantic information and link-level relation information to improve the fine-grained performance of pre-training.
\revision{
Finally, we pre-train RPT on big unlabeled researcher dataset extracted from DBLP, ACM, and MAG digital library.
We set two transfer modes for RPT and fine-tune RPT on three researcher data mining tasks including researcher classification, collaborator prediction, and top-k researcher retrieval to evaluate the effectiveness and transferability of RPT. We also conduct ablation studies and hyper-parameters sensitivity experiments to analyze the underlying mechanism of RPT.
The contributions of our paper are summarised as follows:
}

\revision{
\begin{enumerate}

\item
We introduce the pre-training idea to handle the multiple mining tasks on abundant and heterogeneous researcher data. We propose the RPT framework to extract and transfer useful information from big unlabeled scientific data to benefit researcher data mining tasks.

\item
To perform the RPT framework in considering heterogeneity, generalization, scalability, and transferability, we propose Transformer, GNN based information extractor and a multi-task self-supervised learning objective to capture both the semantic features and local community information of researchers.

\item
Experimental results on real-world researcher data RPT can significantly benefit various downstream tasks. We also conduct model analysis experiments to evaluate different components and hyper-parameters of RPT.
\end{enumerate}
}

\section{Related work}

\subsection{Researcher Data Mining}
\revision{The increasing availability of digital scholarly data offers unprecedented opportunities to explore the structure and evolution of science\cite{Fortunatoeaao0185, lee2017viziometrics,kong2019academic, wan2019aminer}. Multiple data sources such as Google Scholar, Microsoft Academic, ArnetMiner, Scopus, and PubMed cover millions of data points pertaining to researchers(also known as scholars and scientists) community and their output.
Analysis and mining based on big data technology have been implemented on these data and the analyses of researchers have been a hot topic.
The researcher data analysis tasks including collaborator prediction \cite{sun2011co} and recommendation \cite{kong2017exploring,yang2015scientific,nayyeri2020embedding}, collaboration sustainability prediction\cite{wang2019csteller,wang2020scientific}, reviewer recommendation\cite{balachandran2013reducing,zhao2018novel}, expert finding\cite{zhang2007expert,lin2017survey},  advisor-advisee discovery\cite{wang2010mining, zhao2018identifying}, academic influence prediction\cite{kong2017exploring, nie2019academic, dong2016can}, author identification\cite{chen2017task,zhang2018camel} etc.}
Mainstream works focus on mining the various academic characteristics and community graph properties of researchers, then learn task-specific researcher representations for various tasks. For example, here are some recent representative works,
\cite{liu2018context} recommends context-aware collaborator for researchers by exploring the semantic similarity between researchers' published literature and restricted research topics. \cite{wang2019csteller} use researchers' personal properties and network properties as input features to predict the collaboration sustainability. \cite{zhang2020multi} propose an expert finding model for reviewer recommendation, which learn hierarchical representations to express the semantic information of researchers. \cite{liu2019shifu2} propose a network representation learning method on scientific collaboration networks to discover advisor-advisee relationships. \cite{cai2018generative, wang2020collaborative, xia2016scientific} study the problem of citation recommendation for researchers and use the generative adversarial network to integrates network structure and the vertex content into researcher representations. \cite{qiu2018deepinf} incorporate both network structures and researcher features into convolutional neural and attention networks and learn representations for social influence prediction. \cite{yu2020semantic} study the problem of top-k similarity search of researchers on the academic network, the content and meta-path based structure information is embedded into researcher representations.

\subsection{Self-Supervised Learning}
Self-supervised learning is a form of unsupervised learning which aims to train a pretext task where the supervised signals are obtained by data itself automatically, it can guide the learning model to capture the underlying patterns of the data. The key of self-supervised learning is to design the pretext tasks. In the area of computer vision (CV), various self-supervised learning pretext tasks have been widely exploited, such as predicting image rotations\cite{gidaris2018unsupervised}, solving jigsaw puzzles\cite{noroozi2016unsupervised}, and predicting relative patch locations\cite{doersch2015unsupervised}. In natural language processing (NLP), many works propose pretext tasks based on language models, including the context-word prediction\cite{mikolov2013distributed} the Cloze task, the next sentence prediction\cite{kenton2019bert, yang2019xlnet} and so on\cite{chen2019self}. For graph data, the pretext task are usually designed to predict the central nodes given node context\cite{dong2017metapath2vec, you2020graph} or sub-graph context\cite{jiao2020sub}, or maximize mutual information between local and global graph\cite{velickovic2018deep, sun2019infograph}.
Recently, many works\cite{doersch2017multi, ravanelli2020multi, wang2020multi, yang2020html} on different domains has integrated self-supervised learning with multi-task learning, i.e., joint training multiple self-supervised tasks on the underlying models, which can introduce useful information of different facets and improve the generalization performance.

\subsection{Pre-Training Model}
With the idea of self-supervised learning, pre-training models can be applied on big unlabeled data to build more universal representations that work across a wider variety of tasks and datasets\cite{zoph2020rethinking}.
The pre-training models can be classified as feature-based models and end-to-end models.
Early pre-training studies are mainly feature-based, which directly parameterizes the entity embeddings and optimizes them by preserving some similarity measure. The learned embeddings are used as input features, in combination with downstream models to accomplish different tasks. For example,
Word2vec\cite{mikolov2013distributed,mikolov2013efficient}, Glove\cite{pennington2014glove} and  Doc2vec\cite{le2014distributed} in NLP, which are optimized via textual context information.
Early graph pre-training models are similar to NLP, like Deepwalk\cite{perozzi2014deepwalk}, LINE\cite{tang2015line}, node2vec\cite{grover2016node2vec} and metapath2vec\cite{dong2017metapath2vec}, which aim to learn node embeddings to preserve network proximity or graph-based context.
Differently, recent pre-training models pre-train deep neural network-based encoders and fine-tune them along with downstream models in end-to-end manner.
Typical examples includes: MoCo\cite{he2020momentum} and SimCLR\cite{chen2020a} for unlabeled image dataset; BERT\cite{kenton2019bert}, RoBERTa\cite{liu2019roberta} and XLNet\cite{yang2019xlnet} for unlabeled text dataset; GCC\cite{qiu2020gcc} and GPT-GNN\cite{hu2020gpt} for unlabeled graph dataset.
Our proposed RPT is a meaningful attempt to apply a pre-training model on domain-specific and heterogeneous data as the researcher data is scientific data and contains researcher textual attributes and researcher community.

\section{Proposed method}
\subsection{Problem Statement and Framework Overview}
To leverage pre-training models on big unlabeled researcher data, we need to design proper self-supervised tasks to capture the underlying patterns of the data.
The key idea of self-supervised learning is to automatically generate supervisory signals or pseudo labels based on the data itself.
Given the raw data of $N$ researchers, denoted by $A= \{a_1, a_2, ... ,a_N\}$, we first explore the researcher data and extract two categories of researcher features: (1) semantic document set and (2) community graph, for pre-training.

\textbf{Semantic Document Set}.
A researcher may have multiple textual features, including published literature, patents, profiles, curriculum vitae(CV), and personal homepage, and these features may contain rich semantic information with various lengths and different properties. We collect the text information of these features as documents and compose these documents of each researcher together to a organized semantic document set. Formally, the semantic document set of researcher $a_i\in A$ is expressed as $D_i= \{d_1, d_2, ... ,d_{|D_i|}\}$, where every document is formed as a token sequence $d_j= \{t_1, t_2,.., t_{|d_j|}\}$. Noted that each semantic document set $D_i$ and each document $d_j$ may have arbitrary lengths.

\textbf{Community Graph}.
Besides the semantic features, the social communications and relationships are also significant for researcher and have been widely utilized to analysis in previous studies, which can be expressed as graph-structured data.
As the relations between researcher may have different types, We construct the researcher community graph in a heterogeneous graph manner, expressed as $\mathcal{G} = \{(a_i, r_{i,j}, a_j)\} \subseteq A \times \mathcal{R} \times A$, where $(a_i, r_{i,j}, a_j)$ represent one link between researchers, $\mathcal{R}$ is the relation set, and $r_{i,j}\in\mathcal{R}$.
Multiple types of relations between researchers are automatically extracted based on the original data to construct $\mathcal{G}$. For example, we can make the rules that if two researchers coauthor a same paper, there is a relation named $Collaboraing$ between them, if two researchers work for the same organization, there is a relation named $Colleague$ between them. Noted that two researchers can have multiple relations of the same type in $\mathcal{G}$. For instance, if they collaborate on $n$ papers, there would be $n$ $Collaboraing$ relations between them.

\textbf{Researcher Data Pre-Training}.
Formally, the problem of researcher data pre-training  is defined as: given $N$ researchers $A = \{a_1, a_2, ..., a_N\}$, their semantic document sets $\mathcal{D} = \{D_1, D_2,...,D_{N}\}$, and their community graph $\mathcal{G} = \{(a_i, r_{i,j}, a_j)\} \subseteq A \times \mathcal{R} \times A$,
we aim to pre-train a generalized model via self-supervised learning, which is expected to capture both the semantic information and the community information of researchers into low-dimensional representation space, where we hope the researchers with similar researcher topics and academic community are close with each other.
Then, in fine-tuning, the learned model is treated as a generic initialized model for benefiting various downstream researcher mining tasks and is optimized by the task objectives via a few gradient descent steps.
Formally, in the pre-training stages, the pre-training model is expressed as $f_{\theta} =  \Psi(\theta; \mathcal{D}, \mathcal{G})$ and the output is researcher representations $Z = \{\mathbf{z}_1, \mathbf{z}_2,...,\mathbf{z}_N\}$. Let $\mathcal{L}_{pre}$ be the self-supervised loss functions, which extract samples and pseudo-labels in researcher data $\mathcal{D}$ and $\mathcal{G}$ for pre-training. Thus, the objective of pre-training is to optimize the following:

\begin{equation}
\theta_{pre} = \arg \min_{\theta} \mathcal{L}_{pre}(f_{\theta}; \mathcal{D}, \mathcal{G})
\end{equation}

Based on the motivation described above, the pre-training model $f_{\theta}$ optimized by objective $\mathcal{L}_{pre}$ should have the following properties:
\begin{itemize}
    \item \textbf{\textit{Heterogeneity}}. Ability of encoding semantic information from multi-type textual document data and community information from heterogeneous graph data.
    \item \textbf{\textit{Generalization}}.
    Under no supervised information, the pre-training should integrate heterogeneous information  from massive unlabeled data into generalized model parameters and researcher representations.
    \item \textbf{\textit{Scalability}}. As the researcher data is usually massive and contains rich information, the model should be heavy-weight enough to extract the information on the one hand. On the other hand, it needs to be friendly to mini-batch training and parallel computing.
    \item \textbf{\textit{Transferability}}. For fine-tuning on multiple tasks, the model should be compatible with researcher document features and community graph in the downstream tasks.
\end{itemize}

\begin{figure*}
  \centering
  \includegraphics[width=0.99\textwidth]{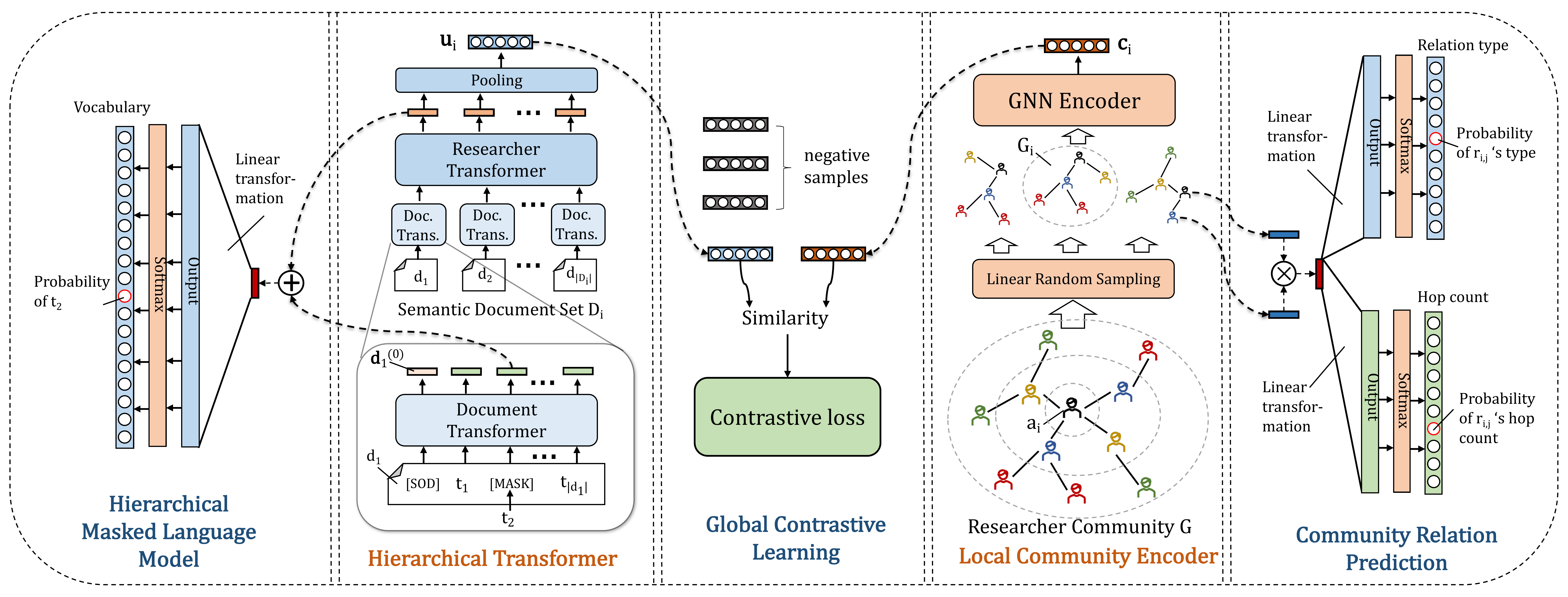}
  \caption{An brief illustration of our proposed researcher data pre-training model. The model contains two encoders, (1) the hierarchical Transformer that encodes the researcher semantic attributes and (2) the local community encoder that encodes the researcher community—noted that in the researcher community, neighbors with different link types to the central researcher are colored differently. Three self-supervised tasks, including global contrastive learning,  hierarchical masked language model, and community relation prediction, are designed to optimize the whole model.}
  \label{framework}
\end{figure*}

As such, the pre-trained model $f_{\theta_{pre}}$ can be adopted on multiple researcher mining tasks. Noted that the focus of this work is on the practicability of the pre-training framework on the researcher data. The goal is to make the model satisfy the above properties, making it different from the common text embedding model or graph embedding model research.
Figure \ref{framework} shows the architecture of the proposed RPT framework.


\subsection{\textbf{Hierarchical Transformer}}
The Transformer model is the state-of-the-art text encoder, which has demonstrated very competitive ability in combination with language pre-training models\cite{kenton2019bert, liu2019roberta}.
Comparing with RNN-like models as text encoder, Transformer is more efficient to learn the long-range dependencies between words. Also, RNN-like models extract text information via sequentially propagation on sequence, which need serial operation of the time complexity of sequence length. While self-attention layers in Transformer connect all positions with constant number of sequentially executed operation. Therefore, Transformer encode text in parallel.
\revision{
A Transformer usually has multiple layers. A layer of Transformer encoder (i.e, a Transformer block) consists of a \textit{Multi-Head Self-Attention Layer}, a \textit{Residual Connections and Normalization Layer}, a \textit{Feed Forward Layer}, and another \textit{Residual Connections and Normalization Layer}. The detailed explanation of the Transformer architecture can be found in \cite{vaswani2017attention}.
Given the input sequence embeddings $X^{(0)} = [\mathbf{x}_1^{(0)}, \mathbf{x}_2^{(0)}, ..., \mathbf{x}_l^{(0)}]$, after the propagation on a  $L$-layers Transformer, formulized as $X^{(L)} = Transformer(X^{(0)})$, we can obtain the final output embeddings of this sequence $X^{(L)} = [\mathbf{x}_1^{(L)}, \mathbf{x}_2^{(L)}, ..., \mathbf{x}_l^{(L)}]$, where each embedding has contained the context information in this sequence. The main hyper-parameters of a Transformer are the number of layers (i.e., Transformer blocks), the number of self-attention heads, and the maximum length of inputs. }

For the semantic document set $D_i= \{d_1, d_2, ... ,d_{|D_i|}\}$ of researcher $a_i$, we aim to use the Transformer to encode the text information in $D_i$ into the researcher representation $\mathbf{u}_i$. Considering that each researcher may have multiple documents with rich text and documents may have different properties and should be processed separately, while the original Transformer can only handle the inputs of a single sentence.
Thus, we propose a two-level hierarchical Transformer model, which consists of a Document Transformer to first encode the text in each document into a document representation, and a Researcher Transformer to integrate multiple documents into the researcher representations.

\textbf{Document Transformer.} For each document $d_j\in D_i$, suppose its token sequence is $\{t_1, t_2,.., t_{|d_j|}\}$ and the corresponding token embeddings are $\{\textbf{t}^{(0)}_1, \textbf{t}^{(0)}_2,.., \textbf{t}^{(0)}_{|d_j|}\}$.
we define a new token named [SOD] (Start Of Document) and random initialize its embedding $\textbf{t}^{(0)}_{[SOD]}$, then we concatenate it with the embedding sequence of $d_j$ into the Document Transformer, which is a multi-layer bidirectional Transformer, and take the final output of [SOD] as the document representation:

\begin{equation}
\begin{aligned}
    \mathbf{d}^{(0)}_j &= [\textbf{t}^{(L_D)}_{[SOD]},\textbf{t}^{(L_D)}_1, \textbf{t}^{(L_D)}_2,.., \textbf{t}^{(L_D)}_{|d_j|}]_{[SOD]} \\
    & = {Transformer_D(\textbf{t}^{(0)}_{[SOD]},\textbf{t}^{(0)}_1, \textbf{t}^{(0)}_2,.., \textbf{t}^{(0)}_{|d_j|})}_{[SOD]}
\end{aligned}
\end{equation}

where $\mathbf{d}^{(0)}_j$ is the $d_j$'s representation, $Transformer_D()$ is the forward function of Document Transformer, $L_D$ is the number of layers, and $[\textbf{t}^{(L_D)}_{[SOD]},\textbf{t}^{(L_D)}_1, \textbf{t}^{(L_D)}_2,.., \textbf{t}^{(L_D)}_{|d_j|}]$ is the output. The input token embeddings are initialized by Word2vec\cite{mikolov2013efficient}, we collect all the texts in the dataset to train the Word2vec model. Also, for each token, its input representation is constructed by summing its embedding with its corresponding position embeddings in documents.



\textbf{Researcher Transformer.} Then, given $a_i$'s semantic document set $D_i =\{d_1, d_2, ... ,d_{|D_i|}\}$, we can obtain the documents' representations $\{\mathbf{d}^{(0)}_1, \mathbf{d}^{(0)}_2, ... ,\mathbf{d}^{(0)}_{|D_i|}\}$ from the Document Transformer and input them into the Researcher Transformer, which is yet another multi-layer bidirectional Transformer but applied on document level, followed with a mean-pooling layer:

\begin{equation}
\begin{aligned}
    \mathbf{u}_i & = Pooling([\mathbf{d}^{(L_R)}_1, \mathbf{d}^{(L_R)}_2,.., \mathbf{d}^{(L_R)}_{|D_i|}]) \\
    & = Pooling
    (Transformer_R(\mathbf{d}^{(0)}_1, \mathbf{d}^{(0)}_2,,.., \mathbf{d}^{(0)}_{|D_i|}))
\end{aligned}
\end{equation}

where $\mathbf{u}_i$ is the semantic representation of researcher $a_i$. $Pooling()$ represent the average of all documents' final outputs. $Transformer_R()$ is the forward function of Researcher Transformer, $L_R$ is the number of layers, and $[\mathbf{d}^{(L_R)}_1, \mathbf{d}^{(L_R)}_2,\mathbf{d}^{(L_R)}_3,.., \mathbf{d}^{(L_R)}_{|D_i|}]$ is the output.
Thus, in this Researcher Transformer, for each researcher's documents set,
each document in the set can collect information from other documents with different attention weights contributed by the self-attention mechanism of the Transformer. So that we can obtain context-aware and researcher-specific document outputs for different researchers, rather than assuming a document (e.g., a paper) has the same contribution to different owners.
Figure \ref{framework} shows the architecture of the proposed hierarchical Transformer, where the Document Transformer is shared by different document inputs and the Researcher Transformer is shared by different researcher inputs.

\revision{
\textbf{Complexity Analysis.}
Instead of splicing each researcher's documents directly into one long text and encoding them using a flat Transformer,
we organize the researcher's text documents in a hierarchical structure and propose the hierarchical Transformer architecture, which significantly improves the efficiency.
Specifically, suppose the number of documents is $n$, the length of each document is $l$, and the hidden dimension is $h$, according to \cite{vaswani2017attention}, the self-attention layer is the most time-consuming component of Transformer and computing the self-attention in the flat Transformer has a quadratic complexity $O((nl)^2h)$ on the total text of researcher's documents. In our approach, the complexity in the document Transformer is $O(l^2h)$ and that in the researcher Transformer is $O(n^2h)$. The total complexity in the hierarchical text encoder is $O(nl^2h + n^2h)$.
Thus, the radio of the computing time of the self-attention in the flat and hierarchical Transformer is $\frac{nl^2}{n+l^2}$.
As usually $n << l^2$, $\frac{nl^2}{n+l^2} \approx n$ and the hierarchical Transformer can be approximately $n$ times faster than the flat Transformer, while their spatial complexity is comparable as the parameter size is independent of the sample length and the number of layers of both would not differ greatly.
}




\subsection{\textbf{Local Community Encoder}}
Given the researcher community graph $\mathcal{G} = \{(a_i, r, a_j)\} \subseteq A \times \mathcal{R} \times A$, first we aim to extract community information of researchers from this graph. Recently, Graph Neural Networks (GNNs) have achieved   state-of-the-art performance on handling graph data \cite{zhang2021hyperbolic, shi2020multi}. Typically, GNNs output node representations via a message-passing mechanism, i.e., they stack $K$ layers to encode node features into low-dimensional representations by aggregating $K$-hop local neighbors' information of nodes.
\revision{
However, taking the whole graph as the input for GNN models can hardly be applied on large-scale graph data due to memory limitations.}
Also, the inter-connected graph structure prevents parallel computing on complete graph topology, making the GNNs propagation on large graph data extremely time-consuming\cite{jiao2020sub}.
One prominent direction for improving the scalability of GNNs is using sub-graph sampling strategies, For example, \revision{instead of using full neighborhood sets, GraphSAGE\cite{hamilton2017inductive} uniformly sample a fixed-size set of neighbors to keep the computational footprint of each batch fixed}. SEAL\cite{zhang2018link} extracts k-hop enclosing sub-graphs to perform link prediction. GraphSAINT\cite{Zeng2020GraphSAINT} propose random walk samplers to construct mini-batches during training.
Thus, we propose to sample a sub-graph of $\mathcal{G}$ to represent the local community graph of researcher $a_i$, denoted by $\mathcal{G}_i$, which can preserve the interactions and relations of $a_i$ with other researchers.
An intuitive way to sample $\mathcal{G}_i$ is to directly sample $a_i$'s $h$-hops neighborhoods, e.g., to sample all $a_i$'s neighbors within $h$-hops as well as corresponding relations to compose $\mathcal{G}_i$.
However, the community graph of researchers usually is denser than other kinds of graphs, each researcher may have dozens of neighbors.
In this sampling way, the size of $\mathcal{G}_i$ might grow geometrically and it would become expensive to process it in training when $h$ increases.

\textbf{Linear Random Sampling.}
In our paper, we propose a linear random sampling strategy,  which can make the number of sampled neighbors increase linearly with the number of sampling hops $h$. The procedure of linear random sampling is presented as follow:

\begin{algorithm}
\caption{Procedure of linear random sampling.}
\KwIn{The researcher $a_i$, the community graph $\mathcal{G}$, the number of sampling hops $h$, and the sampling size $n$.}
\KwOut{The local community graph $\mathcal{G}_i$ of $a_i$.}
$\mathcal{G}^{(1)}_i$ = $SampleOnehopNeighborhood(a_i, n, \mathcal{G})$;\\
\For{$s = 1,.., h-1$}{
    Random sample a $s$-hop neighbor of $a_i$ in $\mathcal{G}^{(s)}_i$, denoted by $a_j$; \\
    $\mathcal{G}^{(1)}_j$ = $SampleOnehopNeighborhood(a_j, n, \mathcal{G})$;\\
    $\mathcal{G}^{(s+1)}_i = \mathcal{G}^{(s)}_i\cup \mathcal{G}^{(1)}_j$;}
\Return $\mathcal{G}^{(h)}_i$; \\
\end{algorithm}

where the $SampleOnehopNeighborhood(a_i, n, G)$ represents the process that randomly sampling $n$ numbers of $a_i$'s 1-hop links, expressed as $(a_i, r, a_j)\in G$, to compose $G^{(1)}_i$, $G^{(s)}_i$ is the sampled local community within $s$-hop.
In this procedure, we obtain a sub-graph of $a_i$'s $h$-hop neighborhood with $n$ neighbors in each hop.

This sampling strategy has the following advantages: (1) The size of local community graphs are linearly correlated to the number of sampling hops, so it would not increase the time complexity of computing community embeddings below when $h$ increases. (2) The sampled neighbor size of each researcher is fixed as $h\times n$ and neighbors with more links are more likely to be sampled.  (3) Noted that we re-sample the local community graphs in each training step, so that all the links in the local community may be sampled after multiple training steps. (4) The sampling strategy can be seen as a data augmentation operation by masking partial neighbors, which is widely used in graph embedding models\cite{jiao2020sub, hamilton2017inductive, velickovic2018deep}. It can help to improve the generalization of models like the mechanism of Dropout\cite{srivastava2014dropout}.



\textbf{GNN Encoder.} Obtained the local community graph $\mathcal{G}_i$ of $a_i$, we use GNN model to encode $\mathcal{G}_i$ into a community embedding. Traditional GNNs can only learn the representations of nodes by aggregating the features of their neighborhood nodes, we refer to these node representations as patch representations. Then, we utilize a Readout function to summarize all the obtained patch representations into a fixed length graph-level representation. Formally, the propagation of $L_C$-layer GNN encoder and the Readout operation is represent as:

\revision{
\begin{equation}
    \mathbf{h}^{(l)}_i = Aggregation^{(l)}(\mathbf{h}^{(l-1)}_i, \{\mathbf{h}^{(l-1)}_j\}: (a_i, r_{i,j}, a_j) \in \mathcal{G}_i)
\end{equation}
\begin{equation}
    \mathbf{c}_i  = Readout(\mathbf{h}^{(L_C)}_j: a_j\in \mathcal{N}(\mathcal{G}_i))
\end{equation}
}

where $0\leq l \leq L_C$, $\mathbf{h}^{(l)}_j$ is the output hidden vector of node $a_j$ at the ${l}$-th layer of GNN and $\mathbf{h}^{(0)}_j$ = $\mathbf{u}_j$, $\widehat{\mathbf{h}^{(l)}_i}$ represents the neighborhood message of $a_i$ passing from all its neighbors in $\mathcal{G}_i$ at the ${l}$-th layer, $Aggregation^{(l)}(\cdot)$ and $Combine^{(l)}(\cdot)$ are component functions of the $l$-th GNN layer, and $\mathcal{N}(\mathcal{G}_i)$ is the node set of $\mathcal{G}_i$. After $L_C$-layer propagation, the output community embedding of $\mathcal{G}_i$ is summarized on node representation vectors through the $Readout(\cdot)$ function, which can be a simple permutation invariant function such as averaging or more sophisticated graph-level pooling function\cite{ying2018hierarchical}.
As the relation between researchers may have multiple types,
in practice,, we choose the classic and widely used RGCN\cite{schlichtkrull2018modeling}, which can be applied on heterogeneous graphs, as the encoder of sub-graphs, we set the layer number $L_C$ same as the sampling hops $h$. Also, we use averaging function for $Readout(\cdot)$ in consider of efficiency.


\revision{
\textbf{Complexity Analysis.}
For simplicity, we omit the relations and discuss the time complexity of applying GCN as the encoder on the researcher community, which can be approximately considered as the time complexity of applying RGCN.
Suppose $m$ is the number of nodes, $d$ is the average node degree in the graph, and $f$ is the hidden dimension.
According to~\cite{NEURIPS2020_a7789ef8}, the time complexity of a $l$-layer GCN on the complete graph can be bounded by $O(lmdf+lmf^2)$, where $O(lmdf)$ is the cost of feature propagation, i.e, the sparse-dense matrix multiplication on the normalized adjacency matrix and the node hidden embedding matrix, and $O(lmf^2)$ is the cost of the feature transformation by applying weight matrix.
Notably, the feature propagation is the dominating complexity term of GCN, and performing propagation on the full neighborhood is the main bottleneck for achieving scalability~\cite{NEURIPS2020_a7789ef8}.
In our local community encoder, the cost of feature transformation is the same. We sample sub-graphs with reduced node degree $n<d$, the cost of feature aggregation to the central nodes is $O(lnf)$, then the per-epoch cost is bounded by $O(lmnf) < O(lmdf)$, which is more efficient than propagation on the complete graph. In addition, sampling and encoding sub-graphs can be easily parallelized, which can further improve efficiency.
}

\textbf{Substitutability of Local Community Encoder.} It is worth mentioning that our method places no constraints on the choices of the local community's sampling strategy and GNN encoder. Our framework is flexible with other neighborhood sampling methods, such as random walk with restart\cite{tong2006fast}, hierarchical tree schema\cite{qiao2020tree}.
Also, traditional GNN such as GCN\cite{kipf2016semi}, GraphSAGE\cite{hamilton2017inductive}, and other GNN models that can encode graph-level representations are available for local community graph encoding, such as graph isomorphism network\cite{xu2018powerful}, DiffPol\cite{ying2018hierarchical} can work in our framework. The design of our framework mainly consider the heterogeneity of researcher community graph and the efficiency as the pre-training dataset is very large, the comparison of different sampling strategies and encoders is not the focus of this paper.

\subsection{\textbf{Multi-Task Self-Supervised Learning}}
In this section, we propose the multi-task self-supervised objective for pre-training, which consists of the main task: global contrastive learning, and two auxiliary self-supervised tasks: hierarchical mask language model and community relation prediction.

\subsubsection{\textbf{Global Contrastive Learning}}
We consider the strong correlation between a researcher's semantic attributes and the local community to design a self-supervised contrastive learning task. The assumption is that given a researcher's local community, we can use the community embedding to infer the semantic information of this researcher, based on the fact that we can usually infer a researcher's research topics according to the community he/she belongs to, and vice versa.
Obtained a researcher embedding $\mathbf{u}_i$ and the embedding $\mathbf{c}_i$ of one sampled local community,
this task aim to discriminate whether they belong to the same researcher. We define the similarity between them as the dot-product of their embeddings and adopt the infoNCE\cite{oord2018representation} loss as our learning objective:

\begin{equation}
\label{eq:cl}
    \mathcal{L}_{Main} = -log\frac{exp(\mathbf{c}_i^T \mathbf{u}_i/\tau)}{exp(\mathbf{c}_i^T \mathbf{u}_i/\tau) + \sum\limits_{a_j\in N_{neg}(a_i)} exp(\mathbf{c}_i^T \mathbf{u}_j/\tau)}
\end{equation}

where $N_{neg}(a_i)$ is the random sampled negative researchers set for $a_i$, its size is fixed as $k$. $\tau$ is the temperature hyper-parameter.
By minimize $\mathcal{L}_{main}$, we can simultaneously optimize the semantic encoder and local community encoder for researchers. Thus, the purpose of the contrastive learning is to preserve the semantic and community information into the model parameters and help the model to integrade these two kinds of information.



\subsubsection{\textbf{Hierarchical Masked Language Model}}
\begin{figure}[]
  \centering
  \includegraphics[width=0.4\textwidth]{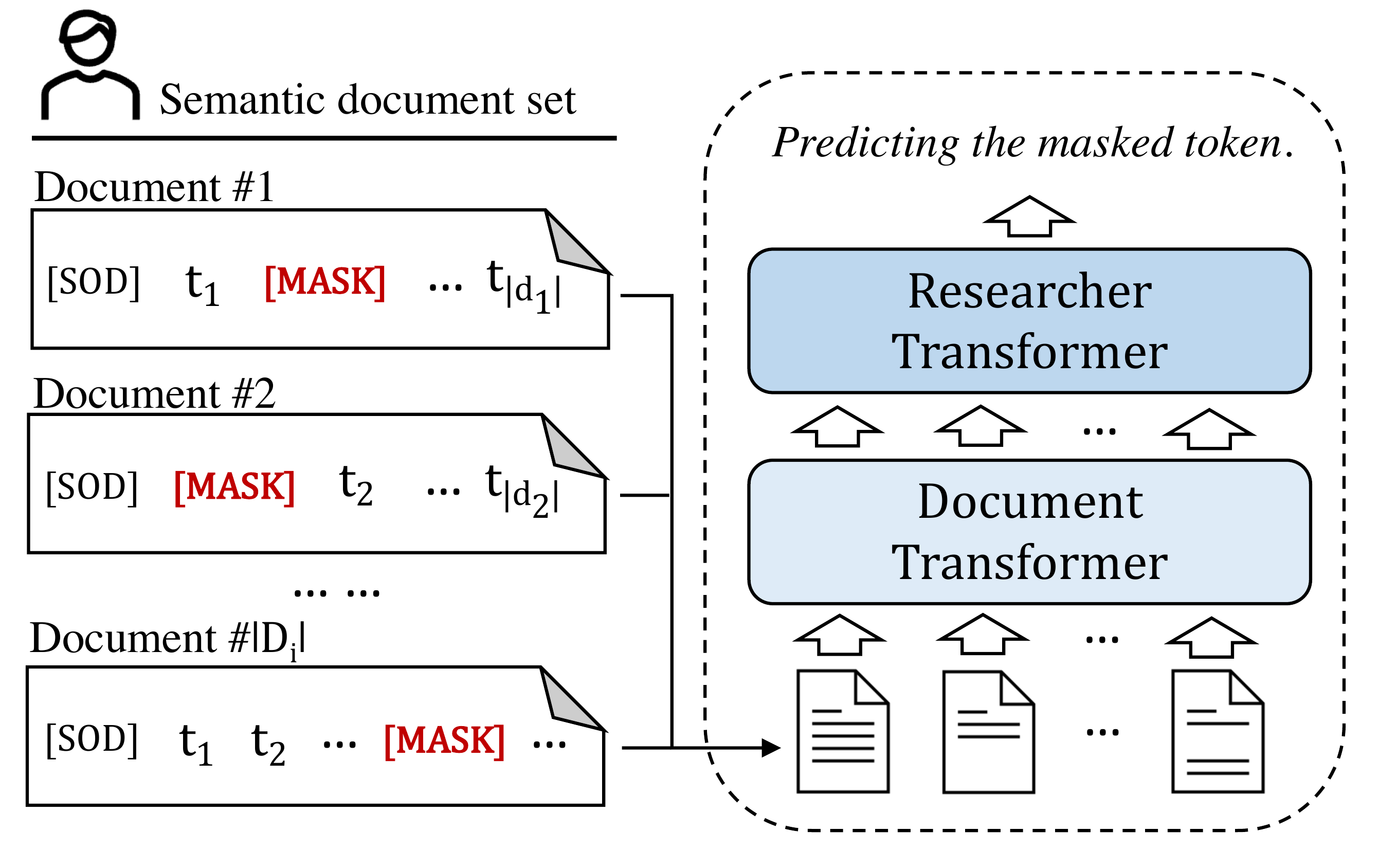}
  \caption{\revision{Hierarchical masked language model. 15\% of the tokens in each document are masked for prediction.}}
  \label{hmlm_pre}
\end{figure}

In the main task, the researcher-level representations are trained via the contrastive learning with their community embeddings. While the document-level and token-level representations are not directly trained.
Inspired by the Mask Language Model(MLM) task in Bert, we propose the Hierarchical Masked Language Model(HMLM) on the hierarchical Transformer to train the hidden outputs of documents and tokens, which can further improve the researcher representations.
As shown in Figure \ref{hmlm_pre}, the MLM task masks a few tokens in each document and use the Transformer outputs corresponding to these tokens, which have captured the context token information, to predict the original tokens.
As our semantic encoder is a two-level hierarchical Transformer,
besides the token-level context captured in the Document Transformer, the Researcher Transformer can capture document-level context (i.e., other documents belong to the same researchers) information, which we assume is also helpful to predict the masked token.
Thus, Given a researcher $a_i$'s semantic document set $D_i= \{d_1, d_2, ... ,d_{|D_i|}\}$, where $d_j\in D_i$ is one document of $a_i$ and the textual sequence of $d_j$ is expressed as $d_j= \{t_1, t_2, ... ,t_{|d_j|}\}$. We first mask 15\% of the tokens in each document. Suppose $t_k\in d_j$ is one masked token and it is replaced with [MASK], the HMLM task aims to predict the original token $t_k$ based on the sequence context in $d_j$ and document context in $D_i$.
First, we obtain the output of the Document Transformer in the position of $t_k$:

\begin{equation}
\begin{aligned}
    \mathbf{t}^{(L_D)}_k &= [\textbf{t}^{(L_D)}_{[SOD]},\textbf{t}^{(L_D)}_1, \textbf{t}^{(L_D)}_2,.., \textbf{t}^{(L_D)}_{|d_j|}]_{t_k} \\
    & = {Transformer_D(\textbf{t}^{(0)}_{[SOD]},\textbf{t}^{(0)}_1, \textbf{t}^{(0)}_2,.., \textbf{t}^{(0)}_{|d_j|})}_{t_k}
\end{aligned}
\end{equation}

where the input embedding $\textbf{t}^{(0)}_k$ of $t_k$ is replaced as the embedding of [MASK]. Then, we obtain the output of Researcher Transformer in the position of document $d_j$, which is the document $t_k$ is in:

\begin{equation}
\begin{aligned}
    \mathbf{d}^{(L_R)}_j & = [\mathbf{d}^{(L_R)}_1, \mathbf{d}^{(L_R)}_2,\mathbf{d}^{(L_R)}_3,.., \mathbf{d}^{(L_R)}_{|D_i|}]_{d_j} \\
    & = Transformer_R(\mathbf{d}^{(0)}_1, \mathbf{d}^{(0)}_2,\mathbf{d}^{(0)}_3,.., \mathbf{d}^{(0)}_{|D_i|})_{d_j}
\end{aligned}
\end{equation}

After that, We sum up these two outputs and fed it into a linear transformation and a softmax function:

\begin{equation}
     \hat{y}_{t_k} = Softmax(W\cdot (\mathbf{t}^{(L_D)}_k + \mathbf{d}^{(L_R)}_j) + b)
\end{equation}

where $\hat{y}_{t_k}\in \mathbb{R}^{V}$ is the probability of $t_k$ over all tokens, $V$ is the vocabulary size. Finally, the HMLM loss can be formulazed as a cross-entropy loss for predicting all masked tokens in $a_i$'s documents:

\begin{equation}
     \mathcal{L}_{HMLM} = -\sum_{d_j\in D_i} \sum_{t_j \in M_{d_j}}\sum_{c=1}^V {y}_{t_k}\ln\hat{y}_{t_k}
\end{equation}

where ${y}_{t_k}$ is the one-hot encoding of $t_k$, $M_{d_j}$ is the set of masked tokens in document $d_j$.


\subsubsection{\textbf{Community Relation Prediction}}
Also, in the main task, we represent the relation information between researchers as the graph-level local community embeddings via linear random sampling and local community encoder. However, the link-level relatedness between researchers is not directly learned. In this auxiliary task, we propose another self-supervised learning task named Community Relation Prediction(CRP), which utilize the links in sampled local communities to construct supervisory signals, as shown in Figure \ref{crp_pre}. The CRP task has two objectives, the first is to predict the relation type between two researchers, the second is to predict the hop counts between two researchers.
Specifically, given a researcher $a_i$'s one sampled local community graph $\mathcal{G}_i$, we random select 15\% links in $\mathcal{G}_i$ as the inputs of CRP task, expressed as $\mathbb{L}^s_i = \{(a_i, l_{i,j}, a_j)\}$, $l_{i,j}$ is the link type between researcher $a_i$ and $a_j$, which can be a atomic relation or composite relations. For example, if $a_i$ and $a_j$ are linked with the relation $r\in \mathcal{R}$, $l_{i,j} = r$; if they are linked by a path $a_i\xrightarrow{r_1}a_k\xrightarrow{r_2}...\xrightarrow{r_m}a_j, r_1,r_2,...,r_m\in \mathcal{R}$, $l_{i,j}$ is the composition from $r_1$ to $r_m$, i.e., $l_{i,j} = r_1\circ r_2\circ,...,\circ r_m$.. In each link $(a_i, l_{i,j}, a_j)$. $l_{i,j}$ has the properties of relation type and hop count, which is what the CRP task aim to predict given the researcher $a_i$ and $a_j$. Thus, we first input the element-wise multiplication of $a_i$ and $a_j$'s output representations into two linear transformations and softmax functions:

\begin{figure}[]
  \centering
  \includegraphics[width=0.4\textwidth]{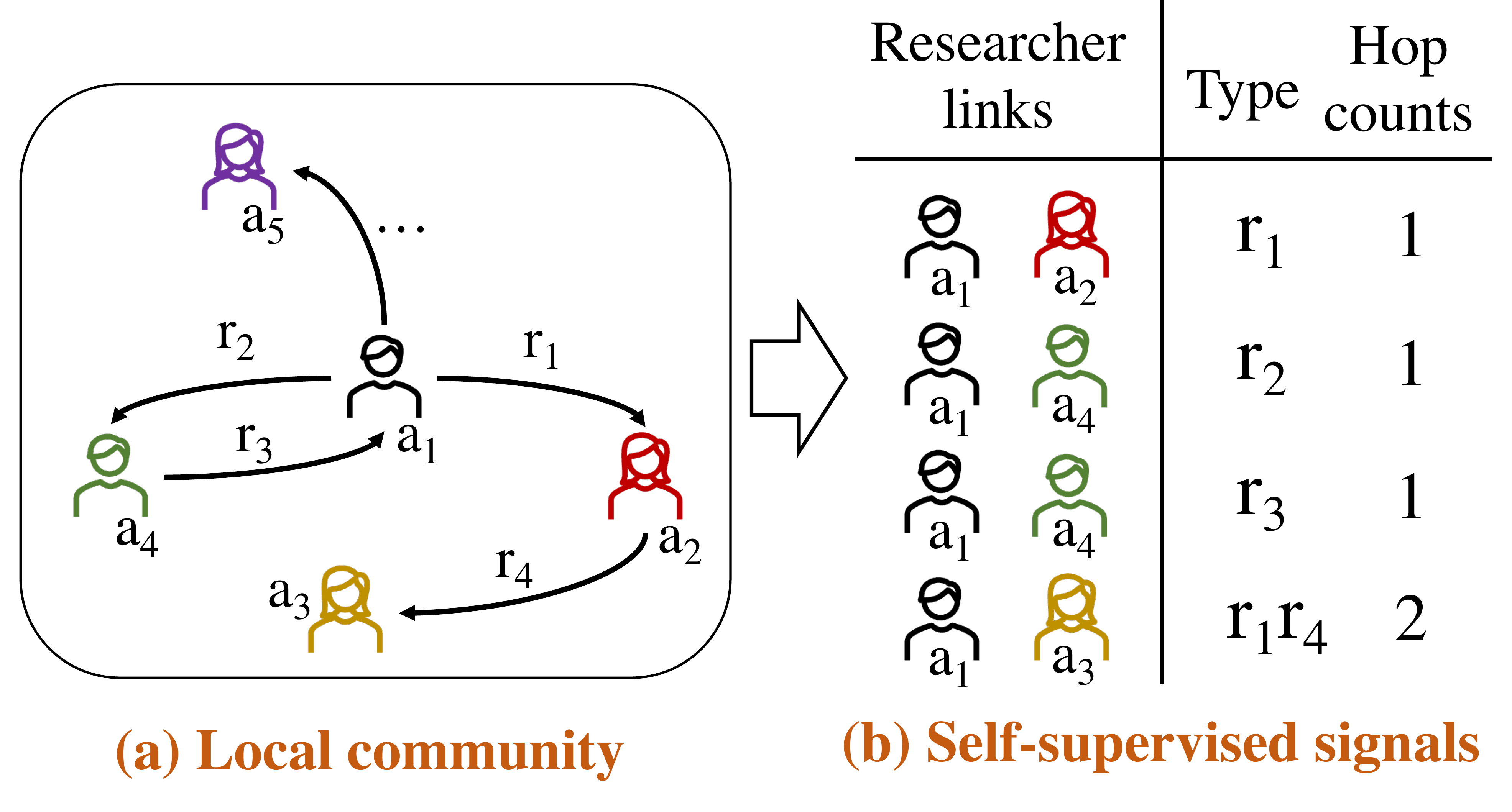}
  \caption{\revision{An illustration of preparing self-supervised signals on the sampled local community of researchers. 15\% links in the community graph is selected for relation prediction.}}
  \label{crp_pre}
\end{figure}

\begin{equation}
     \hat{y}^t_{l_{i,j}} = Softmax(W^t\cdot (\mathbf{h}^{(L_C)}_i \circ \mathbf{h}^{(L_C)}_j) + b^t)
\end{equation}

\begin{equation}
     \hat{y}^h_{l_{i,j}} = Softmax(W^h\cdot (\mathbf{h}^{(L_C)}_i \circ \mathbf{h}^{(L_C)}_j) + b^h)
\end{equation}

where $\hat{y}^t_{l_{i,j}}\in \mathbb{R}^{T}$ is the probabilities of $l_{i,j}$’s type over all link types and $T$ is the number of link types, $\hat{y}^h_{l_{i,j}}\in \mathbb{R}^{H}$ is the probabilities of $l_{i,j}$'s hop count range from 1 to $H$ and $H$ is the maximum hop count from $a_i$ to its neighbors in $\mathcal{G}_i$. Next we input these two probabilities into respective cross-entropy losses to compose the loss function of CRP task:

\begin{equation}
     \mathcal{L}_{CRP} = - \sum_{(a_i, l_{i,j}, a_j)\in \mathbb{L}^s_i}\left( \sum_{c=1}^T {y}^t_{l_{i,j}}\ln\hat{y}^t_{l_{i,j}} +  \sum_{c=1}^H{y}^h_{l_{i,j}}\ln\hat{y}^h_{l_{i,j}}\right)
\end{equation}

where ${y}^t_{l_{i,j}}$ is the one-hot encoding of $l_{i,j}$'s link type index, ${y}^h_{l_{i,j}}$ is the one-hot encoding of $l_{i,j}$'s hop count.
With the guidance of this task, the model can learn fine-grained interaction between researchers such that the GNN encoder is capable of finely capture community information.



\subsection{Pre-training and Fine-tuning}

\textbf{Pre-training.} We leverage a multi-task learning objective by combining the main task and two auxiliary tasks for pre-training.
The final loss function of RPT can be written as:

\begin{equation}
    \mathcal{L}_{pre}(f_{\theta}; \mathcal{D}, \mathcal{G}) = \mathcal{L}_{Main}+ \lambda_1\mathcal{L}_{HMLM} +\lambda_2\mathcal{L}_{CRP}
\end{equation}

where the loss weights $\lambda_1$ and $\lambda_2$ are hyper-parameters. We sample mini-batches of researchers' semantic document sets and local community graphs as the inputs to train the whole model, and
The parameters is optimized by back propagation consistently.

\textbf{Fine-tuning.}
After pre-training, we obtain the pre-trained RPT model $f_{\theta_{pre}}$, which is able to extract valuable information from the researcher semantic document features and community network. Thus, in the fine-tuning stage, given the researcher data $\mathcal{D}^{ft}$ and $\mathcal{G}^{ft}$ for a specific fine-tuning task, we can obtains the semantic representation $\mathbf{u}_i$ and the community representation $\mathbf{c}_i$ of each researcher $a_i$. Suppose $U^{ft}$ and $X^{ft}$ is the semantic and community representation matrix of the fine-tuning task, respectively, the final researcher representation matrix $Z^{ft}$ is obtained from the following:

\begin{equation}
U^{ft}, X^{ft} = f_{\theta_{pre}} =  \Psi(\theta_{pre}; \mathcal{D}^{ft}, \mathcal{G}^{ft})
\end{equation}
\begin{equation}
Z^{ft} = Merge(U^{ft}, X^{ft}, axis=-1)
\end{equation}

where $Merge(\cdot)$ is a merging function to fuse the semantic representation and community representation of researchers, in practice, we directly use the concatenating operation as these two representations have been integrated with each other via contrastive learning in Eq. \ref{eq:cl}. In our framework, we propose two transfer modes of RPT for fine-tuning:

\begin{itemize}
\item
The first is the feature-based mode, expressed as RPT (fb). We treat the RPT as an pre-trained representation generator that first extracts researchers' original features into a low dimensional representation vectors $Z^{ft}$. Then, the encoded representations are used as input initial features of researchers for the downstream tasks.
\item
The second is the end-to-end mode, expressed as RPT (e2e). The pre-trained model $f_{\theta_{pre}}$ with hierarchical Transformer and local community encoder is trained together with each fine-tuning downstream task. Suppose the models of fine-tuning task is $g(\cdot)$ with the parameters $\phi$, the objective of RPT (e2e) can be written as:

\begin{equation}
\theta_{ft}, \phi_{ft} = \arg \min_{\theta,\phi} \mathcal{L}_{ft}(g(f_{\theta_{pre}}, \phi); \mathcal{D}^{ft}, \mathcal{G}^{ft})
\end{equation}

where $\theta_{ft}$ and $\phi_{ft}$ is the optimized parameters, and $\mathcal{L}_{ft}$ is the loss function of the fine-tuning task. All parameters are optimized end-to-end.
\end{itemize}

RPT (e2e) can further extract semantic and community information useful for the downstream researcher data mining tasks. While RPT (fb) without saving and training the pre-trained model is more efficient than RPT (e2e) in the fine-tuning stage.
But compared with traditional solutions training different task-specific models for different tasks from scratch, both these two transfer modes are relatively inexpensive, as the downstream model can be very lightweight and converge quickly with the help of pre-training.

\section{Experiments}
In this section, we first introduce the experimental settings including dataset, baselines, pre-training and fine-tuning parameter setting and implemental hardware and software. Then we fine-tune the pre-trained model on three tasks: researcher classification, collaborator prediction, and top-k researcher retrieval to evaluate the effectiveness and transferability of RPT. Lastly, we perform the ablation studies and hyper-parameters sensitivity experiments to analyze the underlying mechanism of RPT. The code of RPT is publicly available on https://github.com/joe817/RPT. The data in this study is openly available in Science Data Bank at https://datapid.cn/31253.11.sciencedb.01504 and http://www.doi.org/10.11922/sciencedb.01504.

\subsection{Experimental Settings}

\textbf{Dataset.} We perform RPT on the public scientific dataset: Aminer citation dataset\footnote{https://www.aminer.cn/citation}\cite{tang2008arnetminer}, which is extracted from DBLP, ACM, MAG (Microsoft Academic Graph) and contains mullions of publication records.
The available features of researchers contained in the dataset are the published literature, published venues, organizations, etc.
To prepare for the experiments, we select 40281 researchers from Aminer, who have published at least ten papers range from the year of 2013 to 2018.
The information of publication records from 2013 to 2015 are extracted to create the semantic document sets and the community graph of researchers for pre-training.
Specifically, we collect each researcher's papers as his/her semantic documents, the textual sequence of each document is composed by the paper's fields of study. We extract three kinds of relations, \textit{Collaborating} (if they collaborated a paper), \textit{Colleague} (if they are in the same organization), and \textit{CoVenue} (if they published on same venue) between researchers, to construct the researcher community graph (Noted that we random sample 100 neighbors of relation $\textit{CoVenue}$ per researcher). The statistics of semantic document set of researchers and researcher community graph is presented in Table \ref{statistics}.

\begin{table}[htbp]
\centering
\caption{Pre-training Dataset Details.}
\label{statistics}
\begin{tabular}{lc}
\toprule
Number of researchers & 40,281 \\
\midrule
Number of documents  & 225,724 \\
Number of tokens     & 14,391 \\
Average documents per researcher & 13.8 \\
Average tokens per documents & 19.7 \\
\midrule
Number of \textit{Collaborating} & 599,612\\
Number of \textit{Colleague} & 115,891 \\
Number of \textit{CoVenue} & 2,012,935 \\
Average researcher degree of \textit{Collaborating} & 29.8 \\
Average researcher degree of \textit{Colleague} & 5.8 \\
Average researcher degree of \textit{CoVenue} & 100\\
\bottomrule
\end{tabular}
\end{table}


\textbf{Baselines.}
We choose several pre-training models that can capture the information in semantic document sets and researcher community to researcher representations as baselines. Based on if they can be trained end-to-end with the downstream models, we divide these models into feature-based models, including Doc2vec\cite{le2014distributed}, Metapath2vec\cite{dong2017metapath2vec}, and ASNE\cite{liao2018attributed}, and end-to-end models, including BERT\cite{kenton2019bert}, GraphSAGE\cite{hamilton2017inductive}, and RGCN\cite{schlichtkrull2018modeling}.
We also perform our model in feature-based mode and end-to-end mode in fine-tuning.
The detailed descriptions and implementations of baselines are presented as follows:
\begin{itemize}
    \item \textbf{Doc2vec}: Doc2vec is a document embedding model, we collect all the researcher documents from whole dataset to train the model, and use the average embeddings of each researcher's document as the his/her pre-trained embeddings. We use the python gensim library to conduct Doc2vec, we set training algorithm as PV-DM, size of window as 3, number of negtive samples as 5.
    https://pypi.org/project/gensim/.
    \item \textbf{Metapath2vec}: Metapath2vec is a network embedding model, we conduct it on the researcher community graph to obtain node embeddings as pre-trained researcher representations. We set \textit{Collaborating}-\textit{Colleague}-\textit{CoVenue} as the meta-path to sample paths on researcher community via random walk, we set walk length as 10, works per researcher as 5, size of window as 3, number of negative samples as 5.  https://ericdongyx.github.io/metapath2vec/m2v.html.
    \item \textbf{ASNE}: ASNE is a attribute network embedding method which can preserve both the community and semantic attributes of researchers. It concatenates the structure features and node attributes into a multi-layer perceptron to learn node embeddings.
    We use the semantic representations learned by Doc2vec as the input attribute embeddings, and set the the same weight for attribute embedding and structure embedding and the number of hidden layer as 2.
    https://github.com/lizi-git/ASNE.
    \item \textbf{BERT}: BERT is a classic text pre-training model, we concatenate all documents into a sentence as inputs and output the [CLS] output as researcher representations. For a fair comparison, we train the BERT model on our dataset from the beginning. We set the layer of Transformer as 6 and the number of self-attention heads as 8. The maximum length of inputs is set as the same as the product of maximum lengths in Researcher Transformer and Document Transformer in RPT.
    https://github.com/codertimo/BERT-pytorch.
\end{itemize}

We also design two graph pre-training model based on two state-of-the-art GNN models: GraphSAGE and RGCN, GraphSAGE can be applied on homogeneous graph and RGCN can be applied on heterogeneous graphs, and they both can aggregate the local neighborhood information and node attributes into researcher embeddings. we conduct GraphSAGE and RGCN on the researcher community graph and use the self-supervised graph context based loss function introduced in GraphSAGE for pre-training, then in fine-tuning, the pre-trained GraphSAGE and RGCN is trained together with downstream models.

\begin{itemize}
    \item \textbf{GraphSAGE}:
    We use Deep Graph Library tools to build GraphSAGE model. The node attributes in researcher community is initialized by the researcher semantic representations learned by Doc2vec. We use the mean aggregator of GraphSAGE and set the aggregation layer number as 2.
    https://github.com/dmlc/dgl.
    \item \textbf{RGCN}:
    RGCN also used the tools from Deep Graph Library. The node attributes initialization and number of aggregation layers is same with GraphSAGE, we choose the same graph-based loss function from GraphSAGE paper which encourages nearby nodes to have similar representations.
    https://github.com/dmlc/dgl.
\end{itemize}

\revision{
Although some of the above models use both the document information and community information of researchers for pre-training, they may be inclined to extract one kind of information.
Thus, we further design two baselines by model combinations that take advantage of semantic models in extracting textual information and graph-based models in extracting community information.

\begin{itemize}
\item \textbf{D\&M2vec}:A combination of feature-based models--Doc2vec and Metapath2vec. The Doc2vec and Metapathevec is pre-trained based on the above settings to obtain the researcher representations, which are concatenated as the input for fine-tuning tasks.

\item \textbf{BERT\&RGCN}: A combination of end-to-end models--BERT and RGCN.
The BERT and RGCN is first pre-trained based on the above setting. Then, we combine the pre-trained BERT model and RGCN model by concatenating the output embeddings. Both the pre-trained models are co-trained along with downstream models in the fine-tuning tasks.
\end{itemize}
}

\revision{
\textbf{Pre-Training Parameter Setting.}
For the hierarchical Transformer, we set the number of layers of hierarchical Transformer as 6, including three layers of Document Transformer and three layers of Researcher Transformer.
The number of self-attention heads for each Transformer layer is set as 8.
According to the average number of tokens per document and that of documents per researcher, we set the maximum length of inputs as 20 and 10 for Document Transformer and Researcher Transformer, respectively.
For the local community encoder, we set the neighbor sampling hops as 2, the size of sampled neighbor set as 8 for efficiency, and the number of layers of GNN model as 2.
For global contrastive learning, We set the temperature  $\tau$ as 1, and the size of negative samples as 3.
In training, we use Adam optimization with learning rate of 0.01, the exponential decay rates ${\beta}_1 = 0.9$ and ${\beta}_2 = 0.999$, weight decay of 1e-7. we train for 64000 steps with the batch size of 64, which is about 100 training epochs on the pre-training data. We set the researchers representation dimension and all the hidden layer dimensions as 64.
We evaluate different loss weights of two auxiliary tasks and set the optimal values 0.1 for both  $\lambda_1$ and $\lambda_2$.
The code and data is on https://github.com/joe817/RPT.
}

\begin{table}[htbp]
\centering
    \caption{The hyper-parameter settings on three fine-tuning tasks. RC: Researcher Classification; CP: Collaborating Prediction; TRR: Top-k Researcher Retrieval.}
    \label{ftp}
    \begin{tabular}{lccc}
    \toprule
      Parameter & RC & CP & TRR \\
     \midrule
      batch size& 64  & 256 & \ 64 \\
      epochs& 10 & 10 & 20  \\
      dropout rate& 0.9  &  0.9&  0.9 \\
      learning rate& 1e-2  & 1e-3 &  1e-4 \\
      weight decay& 1e-7 & 1e-7 & 1e-4  \\
      adam parameter($\beta_1$) & 0.9 &0.9 & 0.9  \\
      adam parameter($\beta_2$) & 0.999 & 0.999 & 0.999   \\
     \bottomrule
    \end{tabular}
\end{table}

\textbf{Hardware \& Software}
All experiments are conducted with the following setting:
\begin{itemize}
    \item Operating system: CentOS Linux release 7.7.1908
    \item Intel(R) Xeon(R) Gold 6130 CPU @ 2.10GHz.
    \item GPU: 4 GeForce GTX 1080 Ti
    \item Software versions: Python 3.7; Pytorch 1.7.1; Numpy 1.20.0; SciPy 1.6.0; Gensim 3.8.3; scikit-learn 0.24.0
\end{itemize}

\textbf{Fine-tuning Parameter Setting.}
The hyper-parameter settings of RPT on three fine-tuning tasks are presented on Table \ref{ftp}.


\subsection{Task \#1: Researcher Classification}
The classification task is to predict the researcher categories. Researchers are labeled by four areas: \emph{Data Mining (DM), Database (DB), Natural Language Processing (NLP)} and \emph{Computer Vision (CV)}. For each area, we choose three top venues\footnote{\emph{DM}: KDD, ICDM, WSDM. \emph{DB}: SIGMOD, VLDB, ICDE. \emph{NLP}: ACL, EMNLP, NAACL. \emph{CV}: CVPR, ICCV, ECCV.}, then we label researchers by the area with the majority of their publish records on these venues(i.e., their representative areas).
The learned researcher representations by each model are feed into a multi-layer perceptron(MLP) classifier. 
The experimental results are shown in Table \ref{classification}.
The number of extracted labeled authors is 4943, and they are randomly split into the training set, validation set, and testing set with different proportions. We use both Micro-F1 and Macro-F1 as the multi-class classification evaluation metrics.

According to Table \ref{classification}, we can observe that (1) the proposed RPT outperform all baselines by a substantial margin in terms of two metrics, the RPT (e2e) obtains 0.6\%-10.5\% Micro-F1 and 0.7\%-12.4\%  Macro-F1 improvement over baselines.
(2) Noted that the pre-trained model in RPT (fb) is not trained in fine-tuning, comparing four feature-based baselines, RPT (fb) obtains at least 4.5\% improvement, proving that our designed multi-task self-supervised learning objectives can better capture the semantic and community information.
\revision{
(3) Also, both RPT (fb) and RPT (e2e) outperform D\&M2vec and BERT\&RGCN that combines models by embedding concatenation, proving the superior of the global contrastive learning and two auxiliary self-supervised tasks in improving the researcher representations.
}
(4) RPT (e2e) consistently outperforms RPT (fb), indicating the learned parameters in the pre-training model can further contribute to the downstream task by end-to-end fine-tuning.


\begin{table}[htbp]
\caption{Results of Researcher Classification.}
\label{classification}
\scalebox{0.93}{
\begin{tabular}{lcccccc}
\toprule
Proportions(\%)       & \multicolumn{2}{c}{10/10/80} & \multicolumn{2}{c}{20/10/70} & \multicolumn{2}{c}{30/10/60} \\ \midrule
Metric(F1)   &Micro        & Macro      &Micro        & Macro       &Micro        & Macro       \\
\midrule
Doc2vec    &0.781  & 0.759 & 0.799 & 0.780   & 0.804 & 0.787        \\
Metapath2vec &0.760 & 0.733 & 0.769  &0.743  & 0.784 & 0.763  \\
ASNE & 0.790 & 0.770 & 0.811 & 0.786 & 0.810 & 0.797  \\
D\&M2vec & 0.809 & 0.784 & 0.814 & 0.800 & 0.826 & 0.810  \\
\midrule
BERT       & 0.815  & 0.792 & 0.824 & 0.805 &  0.838&  0.825    \\
GraphSAGE    &0.800 &0.777  & 0.812  &0.795 &0.820  &0.808        \\
RGCN        & 0.820 & 0.802 & 0.825 & 0.807 & 0.839 & 0.823     \\
BERT\&RGCN & 0.824 & 0.804 & 0.832 & 0.815 & 0.842 & 0.821  \\
\midrule
RPT (fb) &0.827   &0.805 & 0.848 & 0.833 & 0.852  & 0.837       \\
RPT (e2e)      & \textbf{0.840}  & \textbf{0.824} & \textbf{0.850} & \textbf{0.835}  & \textbf{0.855} & \textbf{0.840} \\
\bottomrule
\end{tabular}}
\end{table}


\subsection{Task \#2: Collaborating Prediction}

Collaborating prediction is a traditional link prediction problem that given the existing collaboration information between authors, we aim to predict whether two researchers will collaborate on a paper in the future, which can be used to recommend potential new collaborators for researchers. To be practical, we randomly sample the collaborating links from 2013 to 2015 for training, in 2016 for validation, and from 2017 to 2018 for testing, noted that duplicated collaborators are removed from evaluation. We use the element-wise multiplication of two candidate researchers' representations as the representation of their collaborating link, then we input the link representation into a binary MLP classifier to predict whether this link exists. Also, negative links (two researchers who did not collaborate in the dataset) with 3 times the number of true links are randomly sampled. We sample various numbers of collaborating links and use accuracy and F1 as evaluation metrics.

The experimental results of different models are reported in Table \ref{collaborator}. According to the table, RPT still performs best in all cases. The following insights can be drawn: (1) Graph representation learning models and GNNs generally achieve better performance than semantic representation learning models, showing that the community information of researchers maybe more important to collaborating prediction.
(2) RPT and RGCN outperform GraphSAGE, indicating the benefit of incorporating heterogeneity information of relations to researcher representations.
\revision{
(3) Incorporating the semantic models with the graph model can further improve the performance. RGCN just uses embeddings of the researcher's documents to initialize node features. While BERT\&RGCN further extracts the document information via multi-layer Transformers, thus achieving better performance.
}
(4) Our methods can achieve 82.9\% accuracy even when the size of the training set is far less than the testing set, indicating the effectiveness of the pre-training model in preserving useful information from unlabeled data.

\begin{table}[htbp]
\caption{Results of Collaborating Prediction. 1k = 1000.}
\label{collaborator}
\scalebox{1}{
\begin{tabular}{lcccccc}
\toprule
Number of links      & \multicolumn{2}{c}{1k/1k/10k} & \multicolumn{2}{c}{3k/1k/10k} & \multicolumn{2}{c}{5k/1k/10k} \\ \midrule
Metric(F1)   &ACC        & F1       &ACC        & F1       &ACC        & F1     \\
\midrule
Doc2vec & 0.764  & 0.240 & 0.767 & 0.310 & 0.778 & 0.363 \\
Metapath2vec &0.778  &0.347  &0.794   &0.391  &0.796  &0.416  \\
AHNE  & 0.796 & 0.483 &0.809  & 0.506 & 0.816 & 0.508 \\
D\&M2vec & 0.796 & 0.389 & 0.797  & 0.421  & 0.809 & 0.449 \\
\midrule
BERT &0.785 & 0.378 & 0.799 & 0.520 & 0.805  &  0.543   \\
GraphSAGE   & 0.777 & 0.308 & 0.781  & 0.477 & 0.795 & 0.482 \\
RGCN         & 0.792 & 0.483 & 0.810 & 0.596 & 0.819 & 0.623  \\
BERT\&RGCN & 0.807 & 0.511 & 0.817 & 0.592 & 0.825 & 0.566 \\
\midrule
RPT (fb)  &0.805   &\textbf{0.624} & 0.827 & 0.622 & 0.828  & 0.633  \\
RPT (e2e)  & \textbf{0.829}  & 0.623 & \textbf{0.840} & \textbf{0.641}  & \textbf{0.860} & \textbf{0.674} \\
\bottomrule
\end{tabular}}
\end{table}

\subsection{Task \#3: Top-K Researcher Retrieval}
The problem of top-K researcher retrieval is a typical information retrieval problem, which is defined as given a researcher, we aim to retrieve several most relevant researchers of him/her.
For each input researcher, we use the dot-product of his and the candidate researcher's representations as their scores and input the scores into a softmax over all researchers to predict top-K relevant researchers, we also use negative sampling\cite{rong2014word2vec} in training for efficiency.
We randomly select 2000 researchers for training, 1000 researchers for validation, and 7000 researchers for testing. The ground truth for each researcher is defined as its coauthor list ordered by the times of collaboration. Noted that we do not introduce any extra parameters except pre-trained models to fine-tuning in this task, and for a fair comparison, the researcher representations are regarded as trainable parameters for feature-based models.
Finally, we use Precision@K and Recall@K in the top-K retrieval list as the evaluation metric and we set K as 1, 5, 10, 15, 20 respectively.

The results are shown in Table \ref{topk}. We can observe that: (1) The performance of feature-based models in this task is far less than end-to-end models in general, comparing their performance in previous tasks. That is because, without downstream parameters, they are inadequate to fit the objective function well. But still, RPT(fb) achieves better performance.
(2) While the end-to-end models can adaptively optimize the parameters in pre-trained models by the downstream objectives, so they can achieve better performance.
(3) The proposed RPT in end-to-end mode still achieves the best performance, showing the designed framework is robust in transferring to different downstream tasks.
\revision{
(4) Experiments on all three tasks show that models fusing both document and community information achieve better results than those inclined to extract one of them, indicating that the researcher's documents and community contain differential information, and both can contribute to the researcher data mining tasks.
}


\begin{table*}[]
\centering
\caption{Results of Top-k Researcher Retrieval}
\label{topk}
\begin{tabular}{lllllllllll}
\toprule
K   & \multicolumn{2}{c}{1} & \multicolumn{2}{c}{5} & \multicolumn{2}{c}{10} & \multicolumn{2}{c}{15} & \multicolumn{2}{c}{20}\\ \midrule
Metric   &Pre@K  &Rec@K  &Pre@K  &Rec@K &Pre@K  &Rec@K &Pre@K  &Rec@K &Pre@K  &Rec@K       \\
\midrule
Doc2vec  &0.278 & 0.041 & 0.152 & 0.099 & 0.107 & 0.131 & 0.088 & 0.154 & 0.076 & 0.171 \\
Metapath2vec & 0.506 & 0.076 & 0.259 & 0.177 & 0.189 & 0.235 & 0.153 & 0.272 & 0.131 & 0.299 \\
ASNE & 0.514   &0.077  &0.288 & 0.192 & 0.198   & 0.244 & 0.159 & 0.279 & 0.134  &0.304 \\
D\&M2vec & 0.580 & 0.077 & 0.326 & 0.185 & 0.216 & 0.227 & 0.167 & 0.253 & 0.138 & 0.271 \\
\midrule
BERT       &0.594    &0.089   &0.299  & 0.205   & 0.195  & 0.246 & 0.149 & 0.272 & 0.124 &  0.291 \\
GraphSAGE    &0.722   & 0.104  &0.389  &0.253  & 0.252  & 0.304 & 0.191 &0.333 & 0.158 & 0.355      \\
RGCN        &0.743 & 0.098  & 0.459  & 0.268  & 0.301 & 0.327 &  0.206    &0.353  & 0.160  & 0.360 \\
BERT\&RGCN & 0.757 & 0.103 & 0.480 & 0.288 & 0.322 & 0.352 & 0.246 & 0.386 & 0.201 & 0.409  \\
\midrule
RPT (fb)      & 0.597  & 0.090  & 0.337 & 0.217  &0.231   &0.273 & 0.182  & 0.309 & 0.154  &  0.337    \\
RPT (e2e)   &\textbf{0.787} & \textbf{0.106}  & \textbf{0.525}  & \textbf{0.309}  & \textbf{0.357} & \textbf{0.381} &  \textbf{0.275}    &\textbf{0.421}  & \textbf{0.226}  & \textbf{0.447} \\
\bottomrule
\end{tabular}
\end{table*}

\subsection{Model Analysis}
In this section, we analyze the underlying mechanism of RPT, we conduct several ablation studies and parameter analysis to investigate the effect of different components, stages, and parameters.

\begin{figure}[]
\centering
\subfigure[Researcher Classification]{
\begin{minipage}[t]{0.48\linewidth}
\centering
\includegraphics[width=1\linewidth]{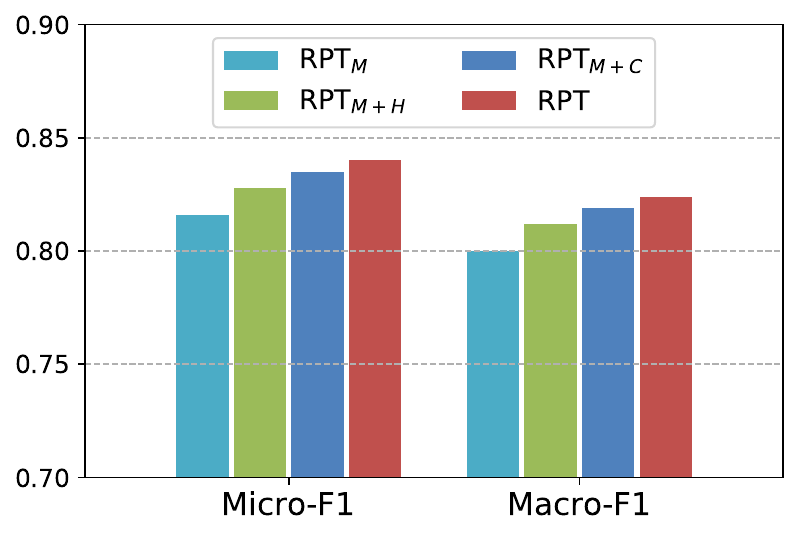}
\end{minipage}%
}%
\subfigure[Collaborating Prediction]{
\begin{minipage}[t]{0.48\linewidth}
\centering
\includegraphics[width=1\linewidth]{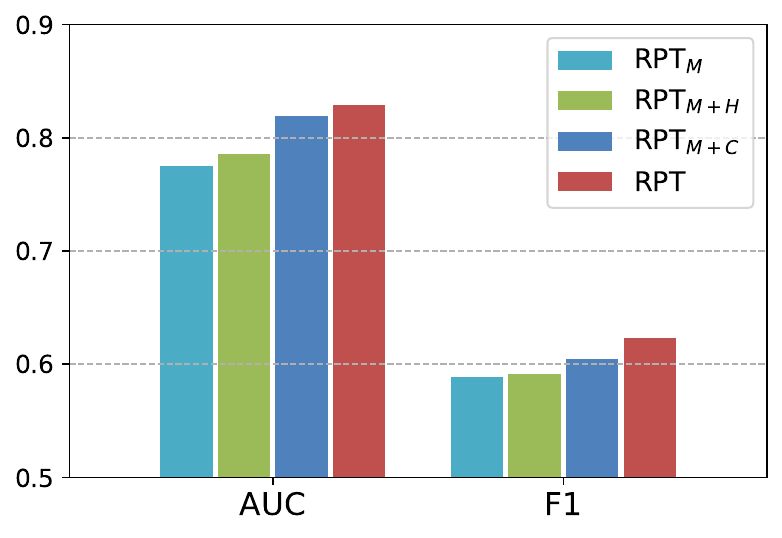}
\end{minipage}%
}%

\caption{Performance evaluation of variant models.}
\label{ablation}
\end{figure}

\begin{figure}[]
\centering
\subfigure[Researcher Classification]{
\begin{minipage}[t]{0.48\linewidth}
\centering
\includegraphics[width=1\linewidth]{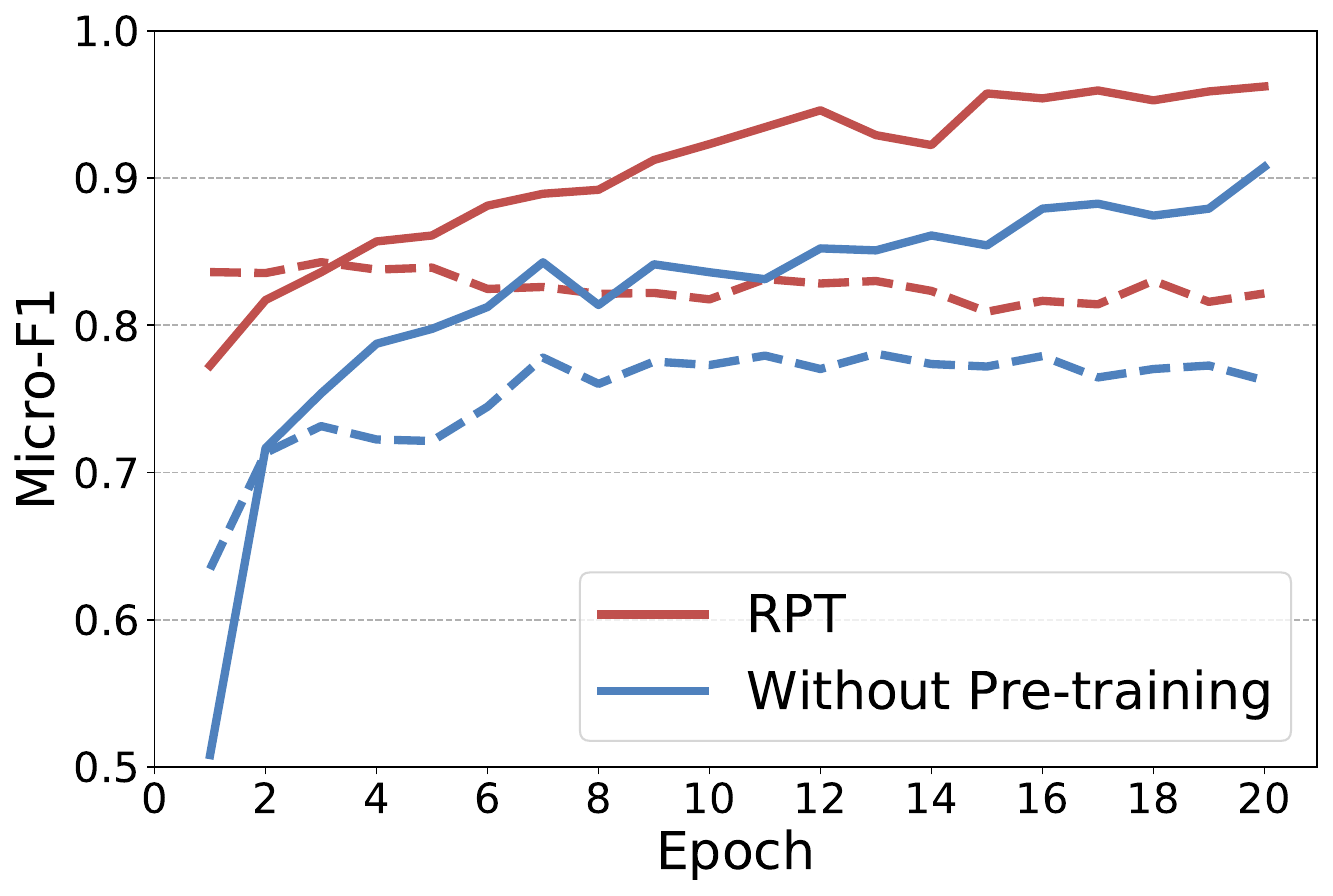}
\end{minipage}%
}%
\subfigure[Collaborating Prediction]{
\begin{minipage}[t]{0.48\linewidth}
\centering
\includegraphics[width=1\linewidth]{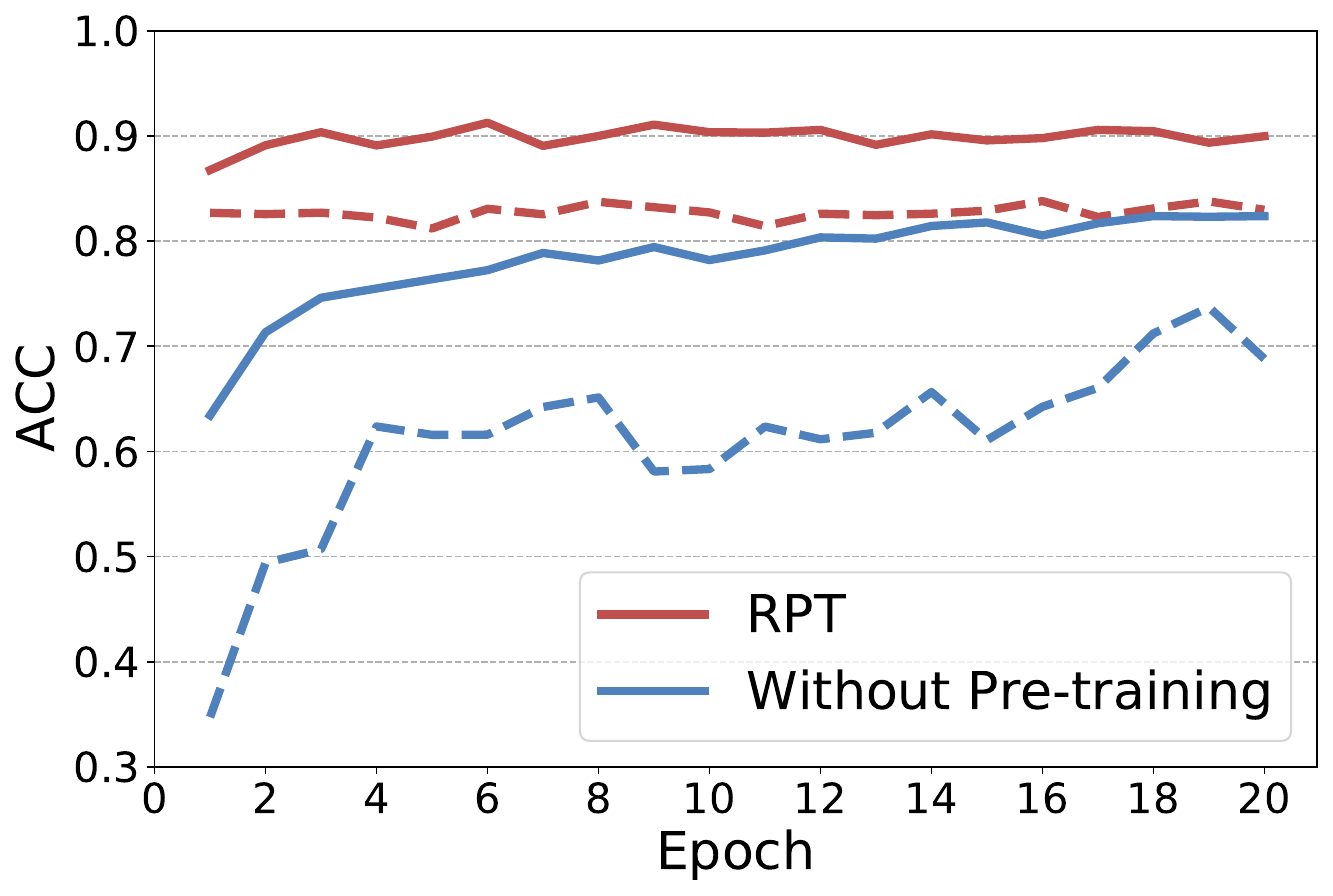}
\end{minipage}%
}%

\caption{Training and testing curves of RPT with pre-training and without pre-training. Solid and
dashed lines indicate training and testing curves, respectively.}
\label{pre-training}
\end{figure}

\textbf{Ablation Study of Multi-tasks.}
As the proposed RPT is a multi-task learning model with the main task and two auxiliary tasks. How different tasks impact the model performance? we propose three model variants to validate the effectiveness of these tasks.

\begin{itemize}
    \item RPT$_{M}$: RPT with only main task $\mathcal{L}_{Main}$.
    \item RPT$_{M+H}$: RPT with main task $\mathcal{L}_{Main}$ and HMLM task $\mathcal{L}_{HMLM}$.
    \item RPT$_{M+C}$: RPT with main task $\mathcal{L}_{Main}$ and CRP task $\mathcal{L}_{CRP}$.
\end{itemize}

We perform these variant models on the researcher classification task and collaborating prediction task, the pre-training setting is same with the complete version, and the fine-tuning is set as the end-to-end mode. Figure \ref{ablation} show the performance of these variants comparing with the original RPT. We can observe that: (1) The results of RPT are consistently better than all the other variants, it is evident that using the three objectives together achieves better performance. (2) Both the RPT$_{M+H}$ and RPT$_{M+C}$ achieve better performance than RPT$_{M}$, indicating the usefulness of both these two auxiliary tasks. (3) RPT$_{M+C}$ is better than RPT$_{M+H}$ on two downstream tasks, which implies
the $\mathcal{L}_{CRP}$ plays a more important role than $\mathcal{L}_{HMLM}$ in this framework.
(4) Comparing the performance of RPT$_{M}$ with the baselines in Table \ref{classification} and \ref{collaborator}, we can find that RPT$_{M}$ still achieves very competitive performance, demonstrating that our framework has outstanding ability in learning researcher representations.

\textbf{Effect of Pre-training.}
To verify if RPT's good performance is due to the pre-training and fine-tuning framework, or only because our designed neural network is powerful in encoding researcher representations. In this experiment, we do not pre-train the designed framework and fully fine-tune it with all the parameters randomly initialized.
In Figure \ref{pre-training}, we present the training and testing curves of RPT with pre-training and without pre-training as the epoch increasing in two downstream tasks.  We can observe that the pre-trained model achieves orders-of-magnitude faster training and validation convergence than the non-pre-trained model. For example in the classification task, it took 10 epoch for the non-pre-trained model to get the 77.9\%  Micro-F1, while it took only 1 epoch for the pre-trained model to get 83.6\% Micro-F1, showing pre-training can improve the training efficiency of downstream tasks. On the other hand, we can observe that the non-pre-trained model is inferior to pre-trained models in the final performance. It proves that our designed pre-training objective can preserve rich information in parameters and provides a better start point for fine-tuning than random initialization.


\begin{figure}[]
\centering
\subfigure{
\begin{minipage}[t]{0.48\linewidth}
\centering
\includegraphics[width=1\linewidth]{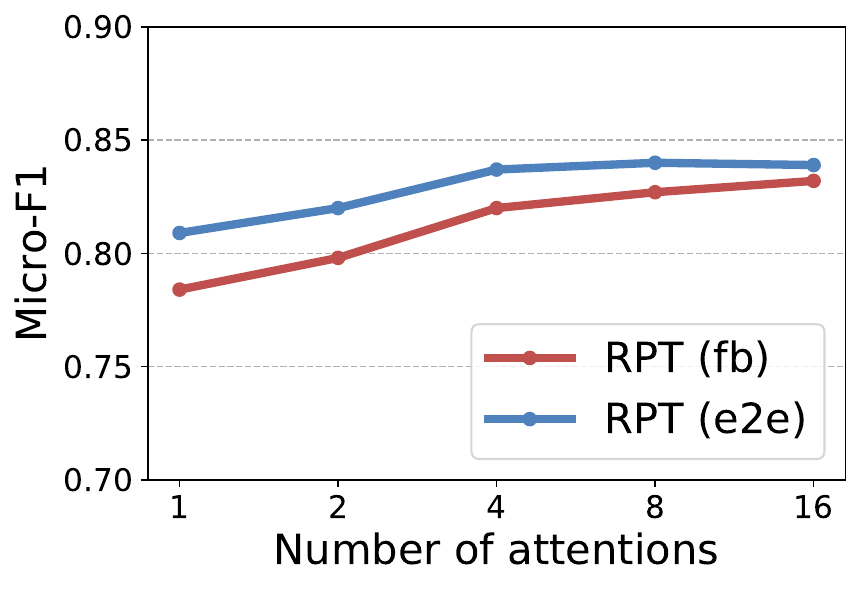}
\end{minipage}%
}%
\subfigure{
\begin{minipage}[t]{0.48\linewidth}
\centering
\includegraphics[width=1\linewidth]{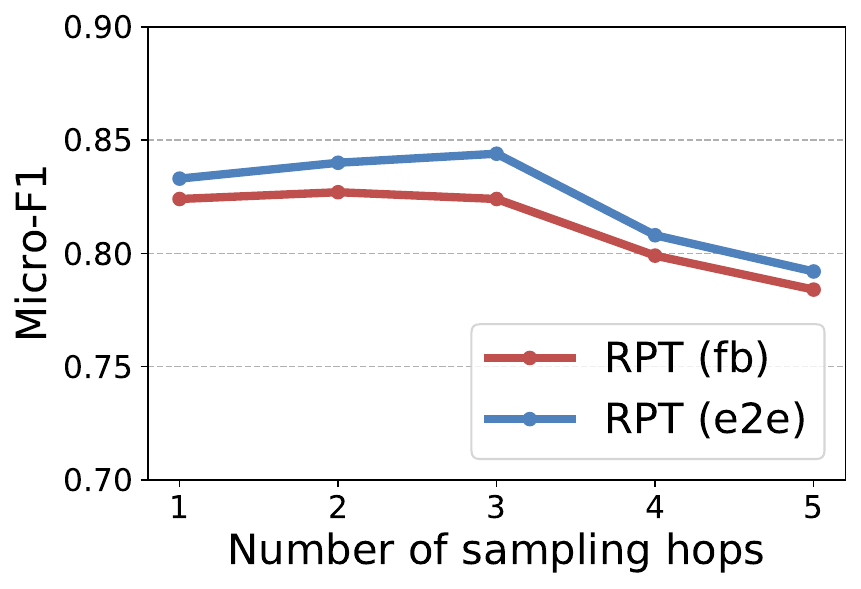}
\end{minipage}%
}%

\caption{Hyper-parameters sensitivity of RPT.}
\label{parameter}
\end{figure}

\textbf{Hyper-parameters sensitivity.}
We also conduct experiments to evaluate the effect of two key hyper-parameters in our model, i.e, the number of self-attention heads in hierarchical Transformers and the sampling hops in the local community encoder. We investigate the sensitivity of these two parameters on the researcher classification task and report the results in Figure \ref{parameter}. According to these figures, we can observe that (1) the more number of attention heads will generally improve the performance of RPT, while with the further increase of attention heads, the improvement becomes slightly. Meanwhile, we also find that more attention heads can make the pre-training more stable. (2) When the number of sampling hops varies from 1 to 5, the performance of RPT increases at first as a suitable amount of neighbors are considered. Then the performance decrease slowly when the hops further increase as more noises (uncorrelated neighbors) are involved.


\section{Conclusion}
In this paper, we propose a researcher data pre-training framework named RPT to solve researcher data mining problems. RPT jointly consider the semantic information and community information of researchers. In the pre-training stage, a multi-task self-supervised learning objective is employed on big unlabeled researcher data for pre-training, while in fine-tuning, we transfer the pre-trained model to multiple downstream tasks with two modes. Experimental results show RPT is robust and can significantly benefit various downstream tasks. In the future, we plan to perform RPT on more meaningful researcher data mining tasks to verify the extensibility of the framework.

\section{Acknowledgments}
This research was supported by the Natural Science Foundation of China under Grant No. 61836013, Ministry of Science and Technology Innovation Methods Special work Project under grant 2019IM020100, Beijing Nova Program of Science and Technology under Grant No. Z191100001119090, Beijing Natural Science Foundation under Grant No.4212030 and Youth Innovation Promotion Association CAS.

\bibliographystyle{IEEEtran}
\bibliography{IEEEabrv,bibfile}


%

\ifCLASSOPTIONcaptionsoff
  \newpage
\fi



%

%

\begin{IEEEbiography}[{\includegraphics[width=1in,height=1.25in,clip,keepaspectratio]{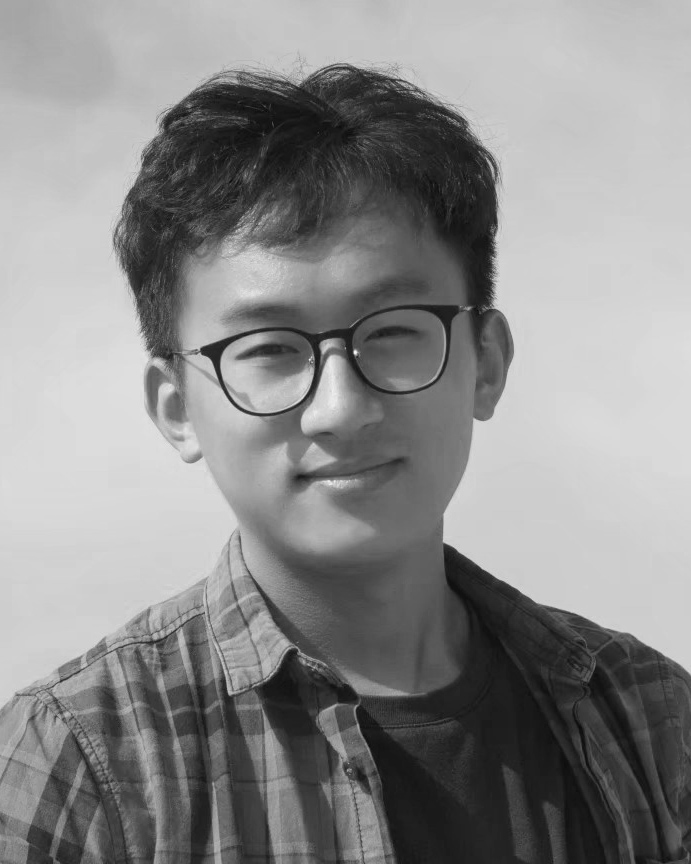}}]{Ziyue Qiao}
was born in 1996. He received his B.S. degree in 2017 from the Wuhan University, China. He is currently working toward obtaining a Ph.D. degree at the Computer Network Information Center, University of Chinese Academy of Sciences. His research interests include networked data mining, graph neural network, graph representation learning, pre-training, and knowledge-based graph.
\end{IEEEbiography}

\begin{IEEEbiography}[{\includegraphics[width=1in,height=1.25in,clip,keepaspectratio]{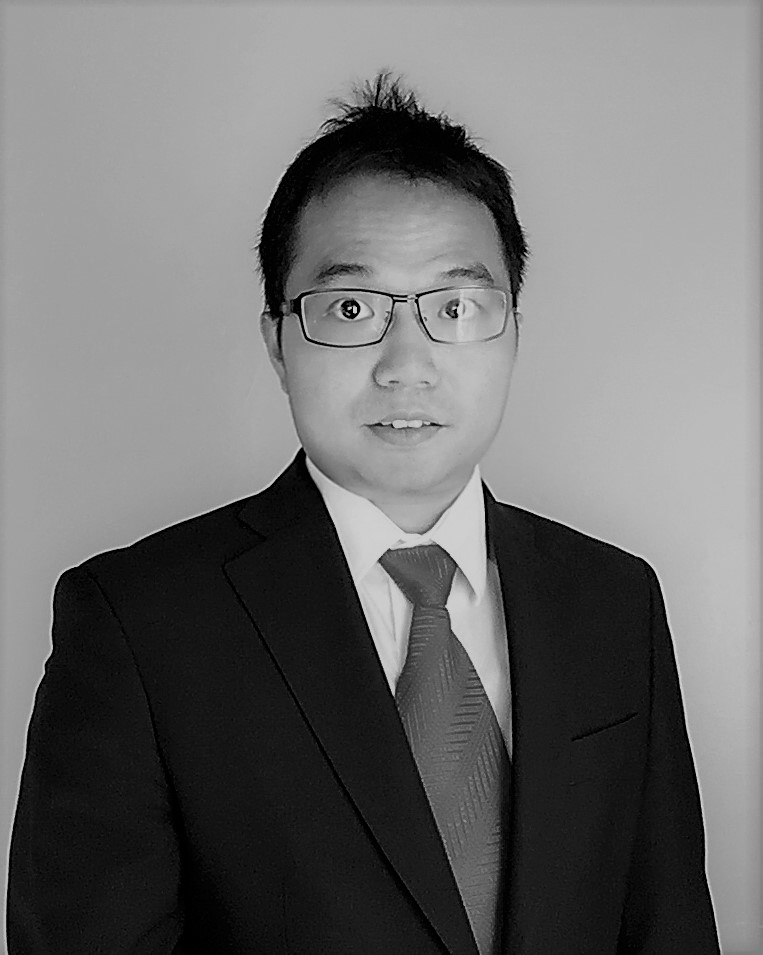}}]{Dr. Yanjie Fu}
is an assistant professor in the Department of Computer Science at the University of Central Florida. He received his Ph.D. degree from Rutgers, the State University of New Jersey in 2016, the B.E. degree from University of Science and Technology of China in 2008, and the M.E. degree from Chinese Academy of Sciences in 2011. His research interests include data mining and big data analytics.
\end{IEEEbiography}

\begin{IEEEbiography}[{\includegraphics[width=1in,height=1.25in,clip,keepaspectratio]{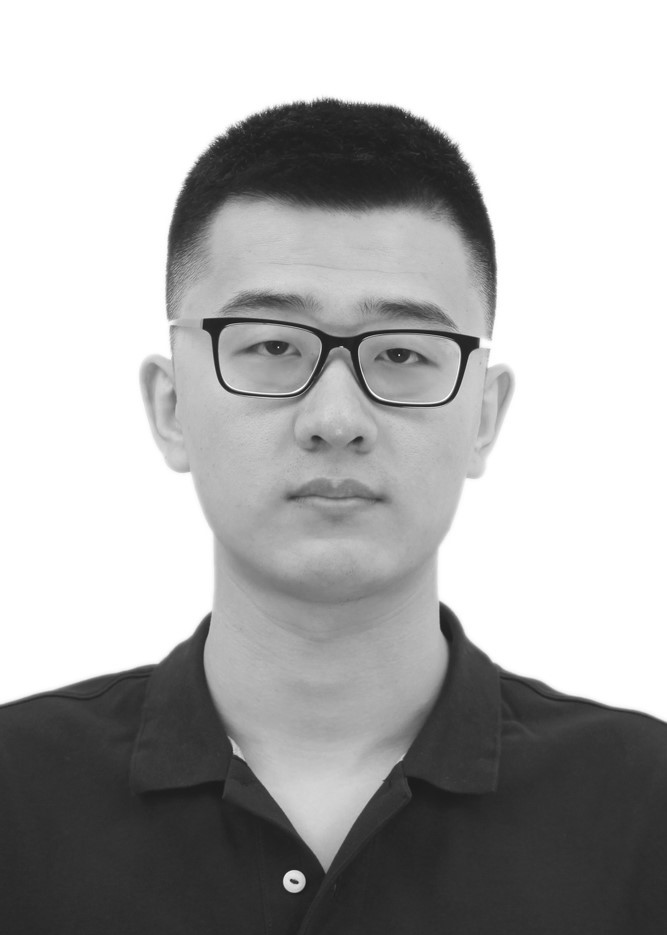}}]{Dr. Pengyang Wang}
is an assistant professor in the State Key Lab of Internet of Things and Smart Cities at the University of Macau. He received his Ph.D. in Computer Science from the University of Central Florida. He has broad interests in data mining and machine learning, especially in geospatial-temporal data modeling. He has published in related domains on top venues such as KDD, TKDE, IJCAI, WWW, AAAI etc. He also served as a (senior) program committee member among top conferences in the data mining and artificial intelligence communities.
\end{IEEEbiography}

\begin{IEEEbiography}[{\includegraphics[width=1in,height=1.25in,clip,keepaspectratio]{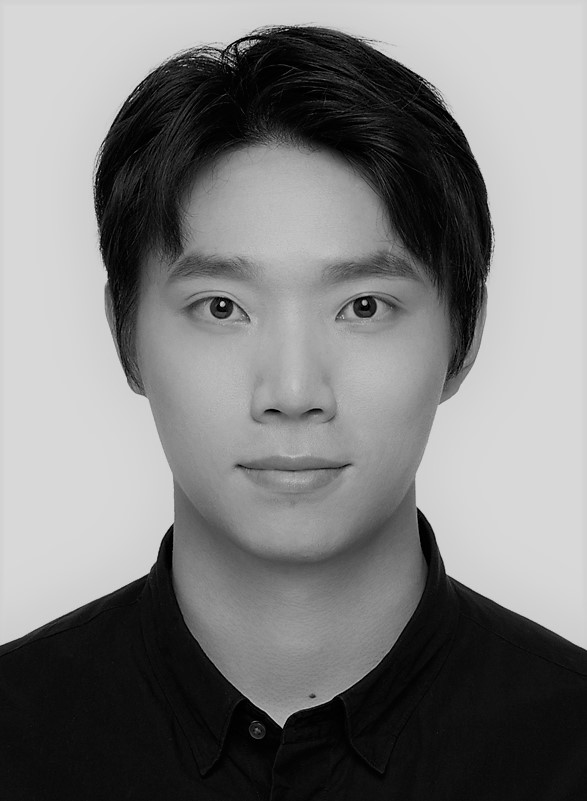}}]{Meng Xiao}
was born in 1995. He received his B.S. degree from the East China University of Technology and graduated from China University of Geosciences (Wuhan) with a master's degree. He is currently working toward obtaining a Ph.D. degree at the University of Chinese Academy of Sciences. His main research interests include Data Mining, Graph Representation Learning, Hierarchical Multi-label Text Classification combined with the knowledge, and the Knowledge Graphs of the discipline system.
\end{IEEEbiography}

\begin{IEEEbiography}[{\includegraphics[width=1in,height=1.25in,clip,keepaspectratio]{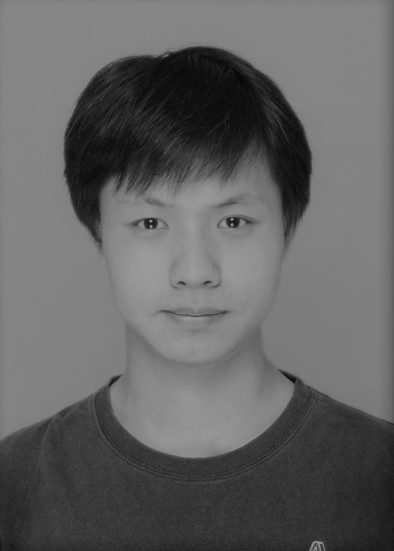}}]{Zhiyuan Ning}
received bachelor degree in 2018 from School of Information Engineering, Zhengzhou University. He is now a PHD candidate in the University of Chinese Academy of Sciences, and is carrying out his research work in Computer Network Information Center of Chinese Academy of Sciences. His research interests focus on natural language processing and knowledge graphs.
\end{IEEEbiography}

\begin{IEEEbiography}[{\includegraphics[width=1in,height=1.25in,clip,keepaspectratio]{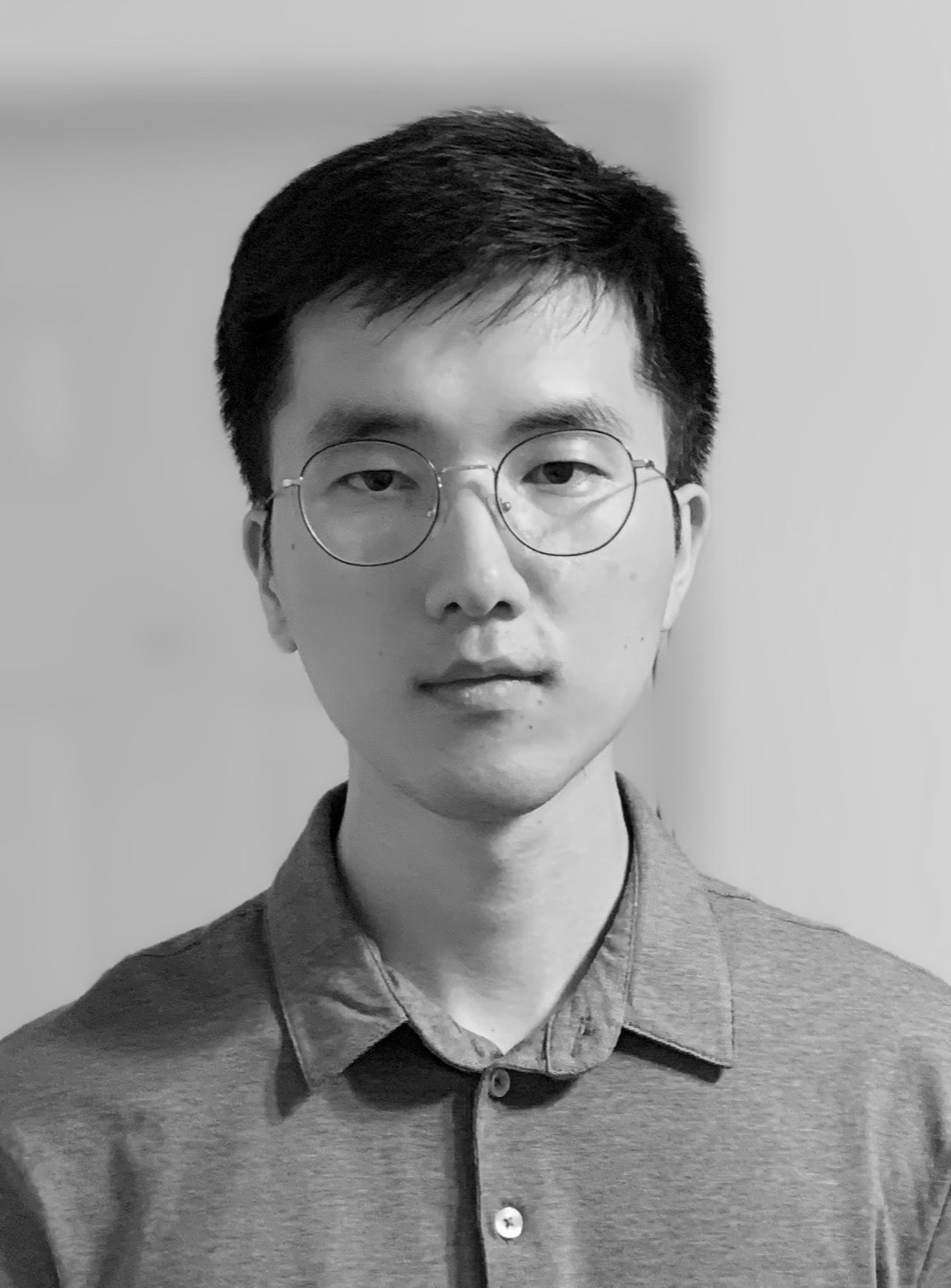}}]{Denghui Zhang}
received the BE degree from
University of Science and Technology Beijing, China, 2015, the MS degree
from Chinese Academy of Sciences, China, 2018.
He is currently working toward the PhD degree in the Information System department at Rutgers University, USA.
His research interests include data mining, business analytics,
natural language processing, and representation learning.
\end{IEEEbiography}

\begin{IEEEbiography}[{\includegraphics[width=1in,height=1.25in,clip,keepaspectratio]{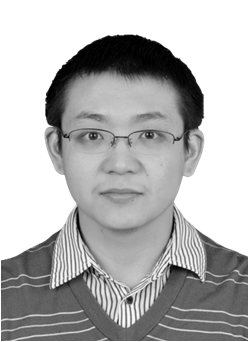}}]{Dr. Yi Du}
was born in 1988. He received his Ph.D. degree from Institute of Software, Chinese Academy of Sciences in 2013. He is an associate professor at the Department of Big Data Technology and Application Development at Computer Network Information Center, Chinese Academy of Sciences. His research interests include big data and visual analytics.
\end{IEEEbiography}

\begin{IEEEbiography}[{\includegraphics[width=1in,height=1.25in,clip,keepaspectratio]{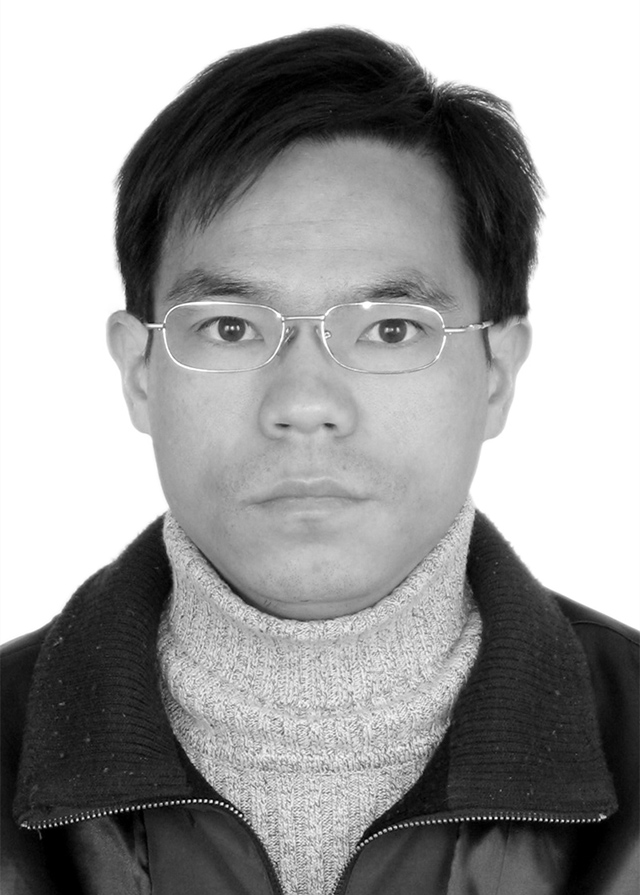}}]{Dr. Yuanchun ZHOU}
was born in 1975. He received his Ph.D. degree from Institute of Computing Technology, Chinese Academy of Sciences, in 2006. He is a professor, Ph.D. supervisor, and the assistant director of Computer Network Information Center, Chinese Academy of Sciences, as well as the director of the Department of Big Data Technology and Application Development. His research interests include data mining, big data processing, and knowledge graph.
\end{IEEEbiography}


\vfill


\end{document}